%% file: paper.tex
\newif \ifdraft    \draftfalse

\documentclass[format=sigconf,10pt,dvipsnames]{acmart}

\usepackage{booktabs} 
\usepackage{verbatim}
\usepackage{hyperref}
\usepackage{paralist}
\usepackage{alltt}
\usepackage{graphicx}
\usepackage[]{fixme}
\usepackage{amsmath}
\usepackage{blkarray}
\usepackage{caption}
\usepackage{subcaption}
\usepackage{multirow}
\usepackage{hhline}
\usepackage{xspace}
\usepackage{algorithm}
\usepackage[noend]{algpseudocode}
\usepackage{setspace}
\usepackage{mathtools}

\usepackage{scalefnt}

\makeatletter
\def\BState{\State\hskip-\ALG@thistlm}
\makeatother

\newtheorem{proposition}{Proposition}

\newcommand{\ScenLoss}{\textit{ScenLoss}}
\newcommand{\AppLoss}{\textit{FlowLoss}}
\newcommand{\AppOptLoss}{\textit{MaxFlowPctLoss}}
\newcommand{\System}{\textit{FloMore}}
\newcommand{\Teavarflow}{\textit{Cvar-Flow-St}}
\newcommand{\Smoreflow}{\textit{Cvar-Flow-Ad}}
\newcommand{\SmoreIP}{\textit{FloMore-Opt}}
\newcommand{\BenderIP}{\textit{FloMore}}
\newcommand{\Smore}{SMORE}
\newcommand{\Teavar}{Teavar}

\newcommand{\flow}{flow}
\newcommand{\MaxCVaR}{\textit{MaxFlowCVaR}}

\newcommand{\SHOWCOMMENT}[1]{\ifdraft {[#1]}\fi}
\newcommand{\sgr}[1]{\SHOWCOMMENT{{\color{red} SGR: #1}}}

\renewcommand\footnotetextcopyrightpermission[1]{} 
\begin{document}
\title{FloMore: Meeting bandwidth requirements of flows}
\author{Chuan Jiang}
\affiliation{%
  \institution{Purdue University}
}
 \email{jiang486@purdue.edu}

\author{Sanjay Rao}
\affiliation{%
  \institution{Purdue University}
}
\email{sanjay@ecn.purdue.edu}

\author{Mohit Tawarmalani}
\affiliation{%
  \institution{Purdue University}
}
  \email{mtawarma@purdue.edu}

\author{\qquad}\affiliation{%
  \institution{\qquad}
}
  \email{ }
\author{Jan 27, 2021}
\affiliation{%
  \institution{\qquad}
}
  \email{}
\author{\qquad }\affiliation{%
  \institution{ \qquad}
}
  \email{ }
\input{abstract}
\maketitle

\input{introduction}

\input{background}
\input{design}

\input{cvar_design}

\input{evaluation}
\input{relatedwork}

\input{conclusion}

\bibliographystyle{plainnat}
\bibliography{sigproc} 

\appendix
\input{appendix}
\end{document}

%% file: abstract.tex
\begin{abstract}
Wide-area cloud provider networks must support the bandwidth requirements of diverse services (e.g., applications, product groups, customers) despite failures. Existing traffic engineering
(TE) schemes operate at much coarser granularity than services, which we show necessitates unduly conservative decisions. To tackle this, we present \System{}, which directly considers the bandwidth needs of individual services and ensures they are met a desired percentage of time. Rather than meet the requirements for all services over the same set of failure states, \System{} exploits a key opportunity that each service could meet its bandwidth 
requirements over a different set of failure states. \System{} consists of an offline phase that identifies the critical failure states of each service, and on failure allocates traffic in a manner that prioritizes those services for which that failure state is critical. We present a novel decomposition scheme to handle \System{}'s offline phase in a tractable manner. Our evaluations show that \System{} outperforms state-of-the-art TE schemes including SMORE and Teavar, and also out-performs extensions of 
these schemes that we devise. The results also show \System{}'s decomposition approach allows it to scale well to larger network topologies.
\end{abstract}

%% file: introduction.tex
\section{Introduction} 
\label{sec:intro}
Cloud providers must ensure that their networks are designed so as to ensure business-critical applications continually operate with acceptable performance~\cite{ciscoWAE,googlesimulator,b4_and_after}. Networks must meet their performance objectives while coping with failure, which are the norm given the  global scale and rapid evolution of networks~\cite{govindan:sigcomm16,failures:sigcomm2011,failures:socc2013,b4_and_after,Sprint,failures:sigcomm2010}.

Networks support multiple services (e.g., applications, product groups, customers), each of which is associated with its own bandwidth requirement, that must typically be met a desired percentage of time. For instance, a customer of a public cloud provider  may indicate that it needs "a bandwidth of at least $B$ between a pair of sites 90\% of the time". Such bandwidth requirements may be expressed at diverse granularities
ranging from individual users to an aggregate set of users~\cite{sigcomm15:BwE}.  Henceforth, in this paper, we use the term \textbf{Flow} to refer to the traffic between a pair of sites corresponding to a particular service, and assume bandwidth requirements are specified at the granularity of \flow{}s. 
Network architects must ensure the requirements of \flow{}s
are met taking into account the likelihood the network may experience different failure states (e.g., a particular set of link failures), and the performance feasible under each failure state.

While bandwidth objectives are specified at the granularity of \flow{}s, existing traffic engineering (TE) schemes manage network resources in a much coarser fashion. First, TE schemes typically compose traffic requirements across all \flow{}s into an aggregate traffic matrix (each cell corresponding to total demand between a pair of sites across all services). Second, TE schemes optimize metrics that relate to the entire aggregate traffic matrix. For instance, many state-of-the-art TE schemes minimize the maximum traffic loss experienced across all pairs of sites~\cite{cvarSigcomm19,pcf_sigcomm20,semi_oblivious_nsdi18}.

The coarse-grained nature of TE schemes makes them unduly conservative. For instance, consider the task of ensuring that all \flow{}s see a desired bandwidth $90\%$ of the time, given the probability the network may encounter different failure states. To achieve this goal, existing TE schemes must ensure that there exist failure states that occur $90\%$ of the time where \textit{all
\flow{}s} must \textit{simultaneously} meet their bandwidth demand. However, operating at the granularity of \flow{}s provides new opportunities. Specifically, each \flow{} could meet its bandwidth requirements in a \textit{different set of failure states}. Thus, the objectives of each \flow{} could be met even though the network may not be able to simultaneously sustain the bandwidth of all \flow{}s $90\%$ of the time.


To tackle this, we present \System{}, a system that considers bandwidth requirements at the granularity of \flow{}s. 
Each flow is associated with a bandwidth demand, and a requirement that the flow's loss must be acceptable in failure states that occur with a desired probability. 

A key idea of \System{} is that not all failure states of the network are critical to meeting a particular flow's requirement. Instead, it seeks to determine the set of \textit{critical failure states} associated with each flow, i.e., those failure states where the loss associated with the flow must be acceptable so the objectives can be met. \System{} consists of (i) an offline phase, which involves determining the critical failure states associated with each flow; and (ii) an online phase (executed when a failure occurs) which determines the bandwidth allocation of all flows recognizing that the failure state is only critical for a subset of flows identified by the offline phase. 

We show that \System{}'s offline phase can be formulated as an Integer Program (IP). The IP however is large, and its complexity grows with network size, the number of failure states of the network that must be considered, and the number of flows. We devise a novel decomposition approach which solves the IP in a series of iterations, where each iteration involves solving much smaller and computationally light-weight optimization problems in a highly parallelizable manner.
The decomposition need not be run to completion, and 
even the initial step, which does not require the solution of any IP, already delivers a better guarantee than that obtained using state-of-the-art TE schemes.

A highlight of \System{} is that it directly optimizes loss at a desired percentile. An alternative is to conservatively estimate this
value using the conditional value at-risk
(CVaR), i.e., the average loss encountered in failure scenarios beyond the desired percentile.
This approach was recently adopted by Teavar~\cite{cvarSigcomm19}, a state-of-the-art TE scheme. However, Teavar operates over coarse-grained traffic matrices, and assumes a less flexible routing strategy 
To facilitate comparisons, we devise new CVaR schemes that operate at the granularity of flows, and consider adaptive routing strategies. 


We evaluate \System{} over 20 topologies including many large topologies. We compare \System{} with Teavar, and SMORE (another state-of-the-art TE schemes). We show that \System{} guarantees much lower loss at a desired percentile relative to SMore and Teavar. Further, we show that while our enhanced CVaR schemes greatly improve performance over Teavar, \System{} still out-performs them. This is 
because CVaR is a weak approximation of the performance at a desired percentile. Our results also show the importance of our decomposition scheme and that it performs close to the optimal with acceptable computation time.


%% file: background.tex
\section{Motivation}
\label{sec:motivation}

\begin{figure}[t]
	\centering
    	    \includegraphics[trim={7.5cm 14.3cm 10.8cm 4.2cm}, clip, width=0.48\textwidth]{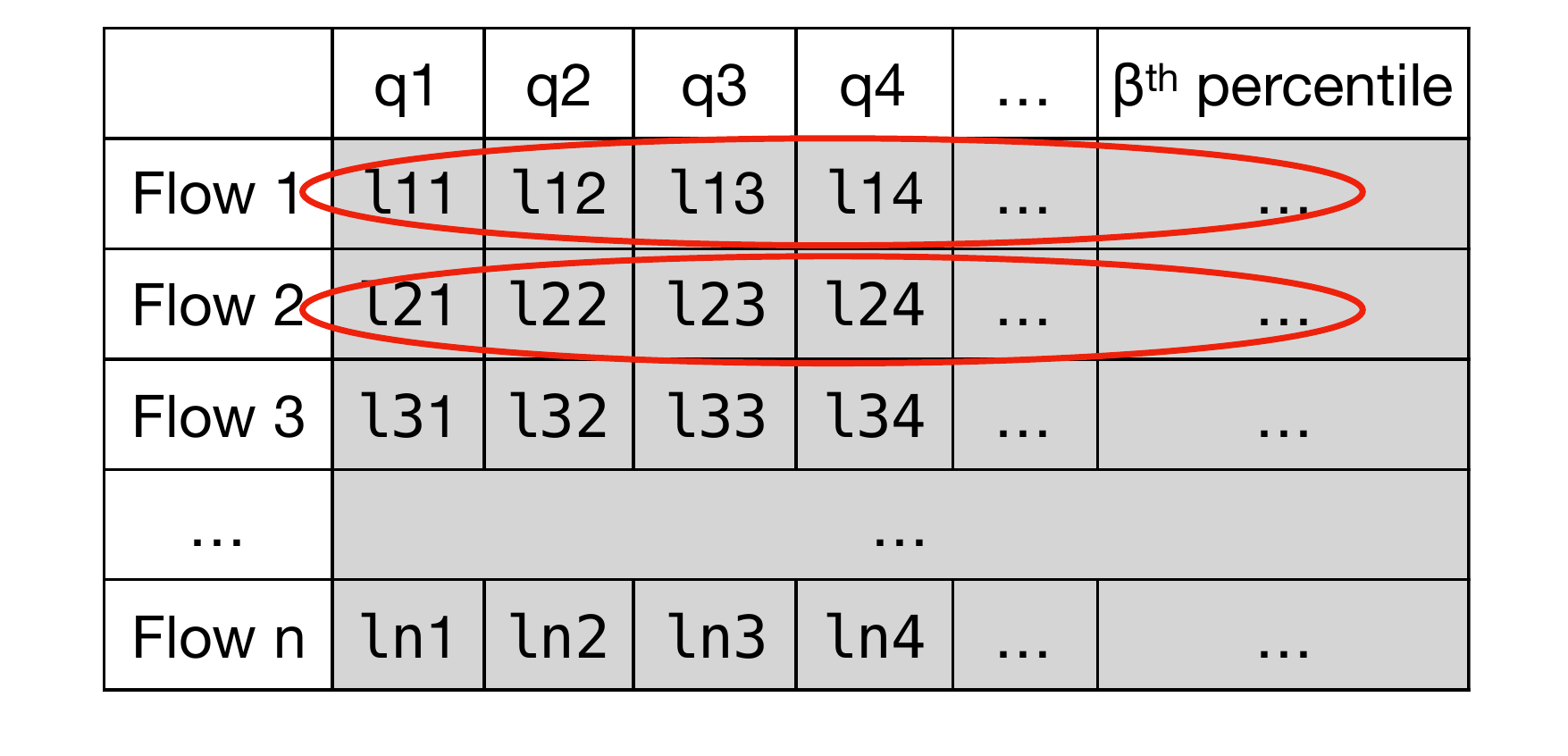}
        \label{fig:app_loss}
\vspace{-0.3in}
    \caption{Meeting bandwidth requirements requires computing the $\beta^{th}$ percentile of flow losses.
    }
        \label{fig:app_loss}
        \vspace{-0.25in}
\end{figure}

\begin{figure}[t]
	\centering
    	    \includegraphics[trim={6.1cm 10.6cm 11.9cm 6.3cm}, clip, width=0.48\textwidth]{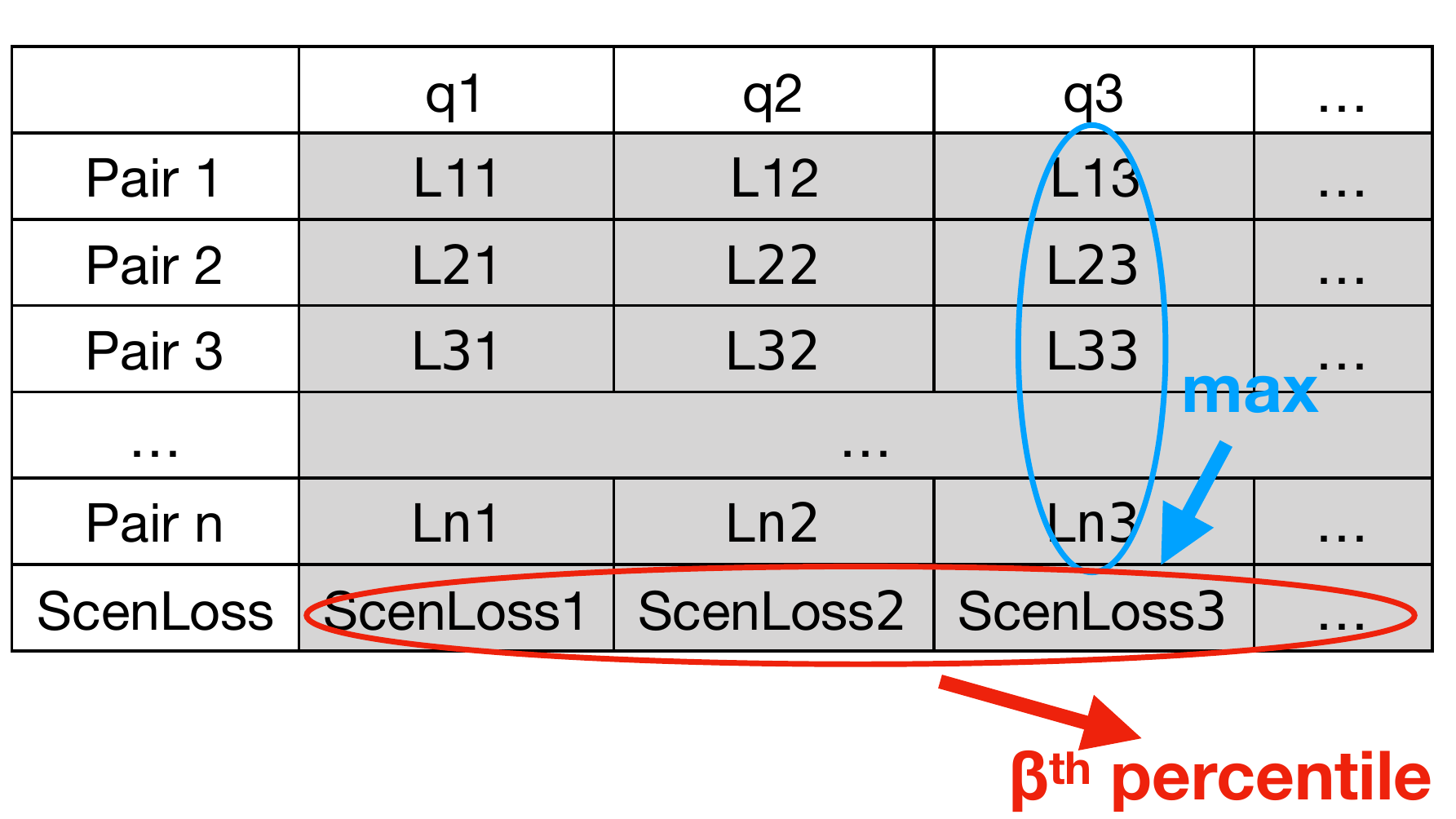}
        \label{fig:pair_loss}
\vspace{-0.25in}
 \caption{
 Existing TE schemes focus on $\beta^{th}$ percentile of \ScenLoss{}
 }
    \label{fig:pair_loss}
    \vspace{-0.2in}
\end{figure}

\subsection{Background and problem context}

\paragraph{How bandwidth objectives are defined.}
A network supports multiple \textit{services}, where each service comprises of multiple \textit{\flow{}s}, and each \flow{} is associated with a particular ingress and egress node. A \flow{} itself could represent traffic at different granularities -- e.g., the traffic of an individual user, an application, an aggregate group of users, a product group, or a customer of a public cloud provider~\cite{sigcomm15:BwE}. 

A \flow{}'s bandwidth requirements must be met despite network failures. Network operators typically have empirical data which indicate the probability of the network experiencing different node/link failures~\cite{cvarSigcomm19}.  We use the terms \textit{failure state} and \textit{failure scenarios} to refer to a state of the network where a particular set of links or nodes have failed. 

Let $q$ denote a failure scenario. Each \flow{} $f$ is associated with a bandwidth requirement $d_f$. 
Since the entire requirement cannot always be met under failure, let $l_{fq}$ denote the 
\textit{bandwidth loss}, i.e., the fraction of the bandwidth requirement $d_f$ that cannot be met 
in scenario $q$. Equivalently, the flow is allocated a bandwidth of $(1 - l_{fq}) d_f$
in scenario $q$.

A \flow{}'s requirements are typically specified as (i) a bandwidth demand $d_f$; and (ii) 
a requirement that despite failures, the flow must see a bandwidth loss of at most $l$ at 
least $\beta$ of the time. 
Consider Fig.~\ref{fig:app_loss}, where each row corresponds to a \flow{}, each column to a 
failure scenario, and each cell shows the 
corresponding loss $l_{fq}$.
To meet \flow{} level requirements, an architect must compute the $\beta^{th}$ percentile
of each row, and ensure the resulting value is at most $l$.

\paragraph{Abstractions and metrics of existing TE schemes.}
Traditionally, traffic requirements are abstracted as a set of source and destination pairs, and the 
traffic demand associated with each pair $i$. Effectively, the traffic requirements of all underlying 
services and their constituent \flow{}s are composed to obtain a traffic matrix. Given such a matrix, 
TE schemes allocate bandwidth to each pair $i$ so a desired performance metric over the entire traffic 
matrix aggregate is optimized, while ensuring link capacities are not exceeded. 

For concreteness, consider Teavar~\cite{cvarSigcomm19}, a recent and state-of-the-art TE scheme. Similar to the 
discussion above, let $L_{iq}$ denote the bandwidth loss for pair $i$ in scenario $q$.
We define the $\ScenLoss{}_q$ as the maximum loss across all source destination pairs in a given scenario.
That is, 
\begin{align}
\ScenLoss{}_q = max_{i} L_{iq} \nonumber
\end{align}
Teavar~\cite{cvarSigcomm19} seeks to allocate bandwidth to each pair so that the worst \ScenLoss{} across a set of scenarios 
that occur with a probability of $\beta$ is minimized.

Pictorially, consider Figure~\ref{fig:pair_loss} which is similar to Figure~\ref{fig:app_loss}, except that
the rows are now source destination pairs, and each cell corresponds to the loss seen by a particular pair 
in a given failure scenario. Teavar effectively computes $\ScenLoss{}_q$ for each scenario $q$ by aggregating
along each column, and then seeks to estimate the $\beta^{th}$ percentile of $\ScenLoss{}_q$ across failure
scenarios. In fact, Teavar conservatively estimates the $beta^{th}$ percentile as we will discuss in \S\ref{sec:cvar_design}.


%

Many other TE schemes also essentially minimize \ScenLoss{}. For instance, many TE schemes~\cite{semi_oblivious_nsdi18,r3:sigcomm10}
including the state-of-the-art SMORE~\cite{semi_oblivious_nsdi18} optimize the utilization of the most congested link 
(Maximum Link Utilization or MLU). Other schemes~\cite{pcf_sigcomm20,ffc_sigcomm14} solve the maximum 
concurrent flow, and maximize the fraction $z$ (that we also refer to as scale factor) of demand the network 
can handle. Minimizing MLU, or maximizing $z$ is equivalent to minimizing \ScenLoss{}, since $\ScenLoss{} = \max\{0,1-z\}$, and $\ScenLoss{} = \max\{0, 1-1/\text{MLU}\}$. 
While most of these other TE schemes do not address percentile requirements,
the natural approach to analyzing their performance under failures is also to compute the $\beta^{th}$ percentile of $\ScenLoss{}_q$ (last row) across failure scenarios similar to the above.


\begin{figure}[t]
	\centering
    	   \includegraphics[trim={6cm 9cm 6.6cm 6.5cm}, clip, width=0.48\textwidth]{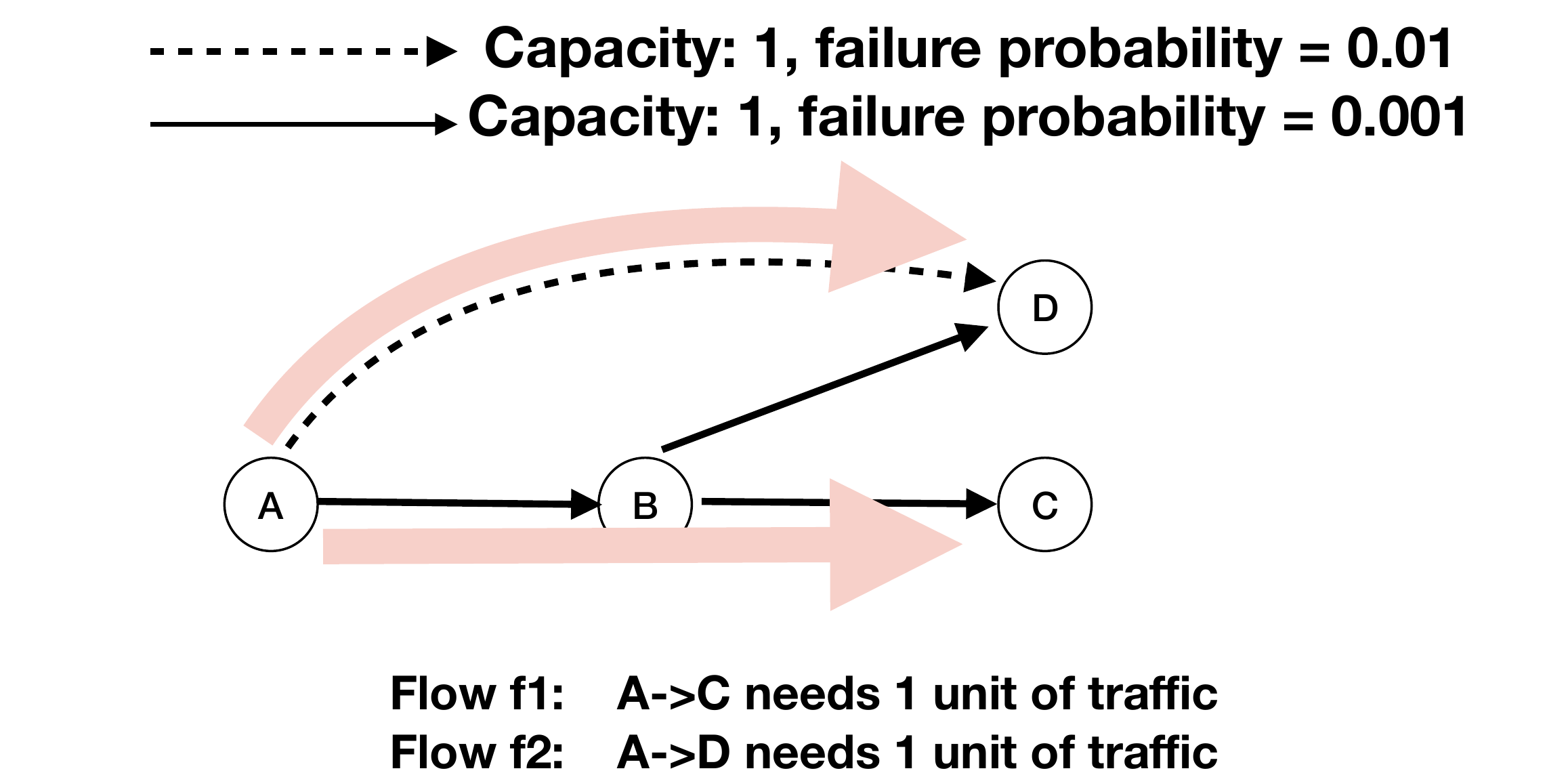}
        \label{fig:flow_example}

    \vspace{-0.25in}
    \caption{Illustrating \System{}’s potential
    }
        \label{fig:flow_example}
        \vspace{-0.1in}
\end{figure}

\begin{figure}[t]
	\centering
    	  \includegraphics[trim={6.7cm 9cm 4.3cm 5cm}, clip, width=0.48\textwidth]{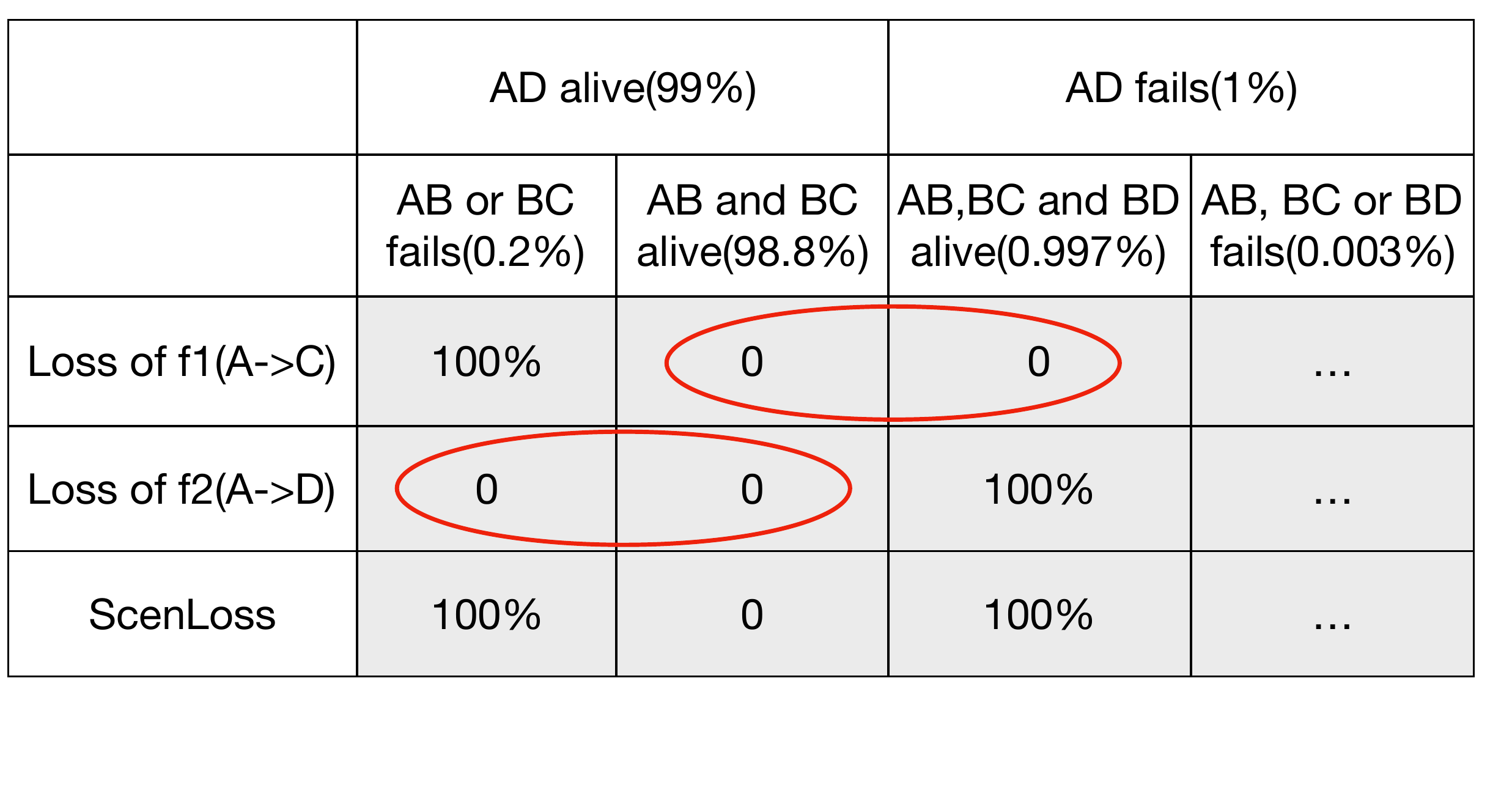}
        \label{fig:optimal_table}
\vspace{-0.4in}
    \caption{The loss achieved for routing scheme in Fig.~\ref{fig:flow_example}. The ovals show that each flow can
   be sent with 99\%ile loss of 0. The last column's values are not relevant.
    }
        \label{fig:optimal_table}
        \vspace{-0.1in}
\end{figure}

\begin{figure}[t]
	\centering
        \includegraphics[ trim={6.8cm 10.5cm 6.7cm 6.5cm}, clip, width=0.48\textwidth]{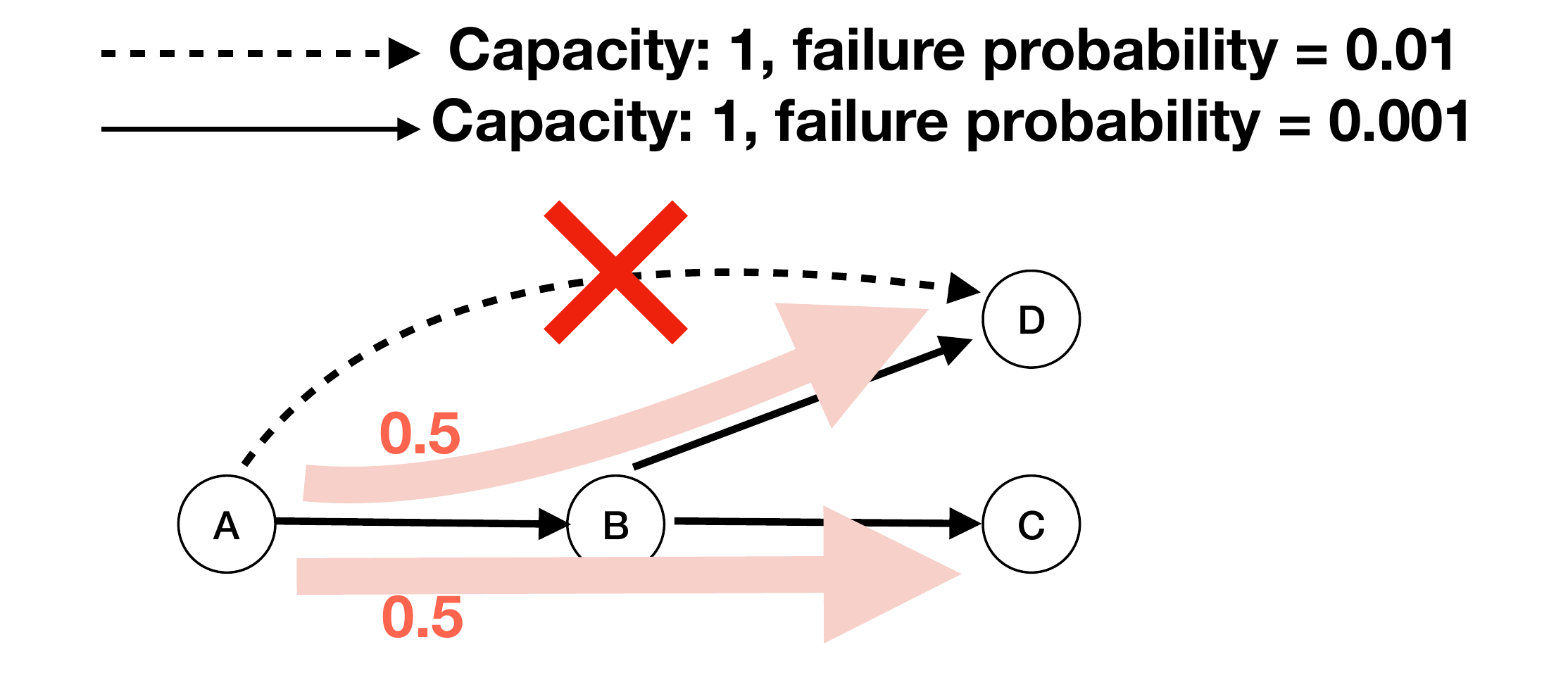}
        \label{fig:flow_example_1}

    \vspace{-0.25in}
    \caption{How traffic is routed by SMORE and Teavar when link AD fails.}
        \label{fig:flow_example_1}
        \vspace{-0.1in}
\end{figure}

\begin{figure}[t]
	\centering
 \includegraphics[trim={6.5cm 9cm 4.3cm 5cm}, clip, width=0.48\textwidth]{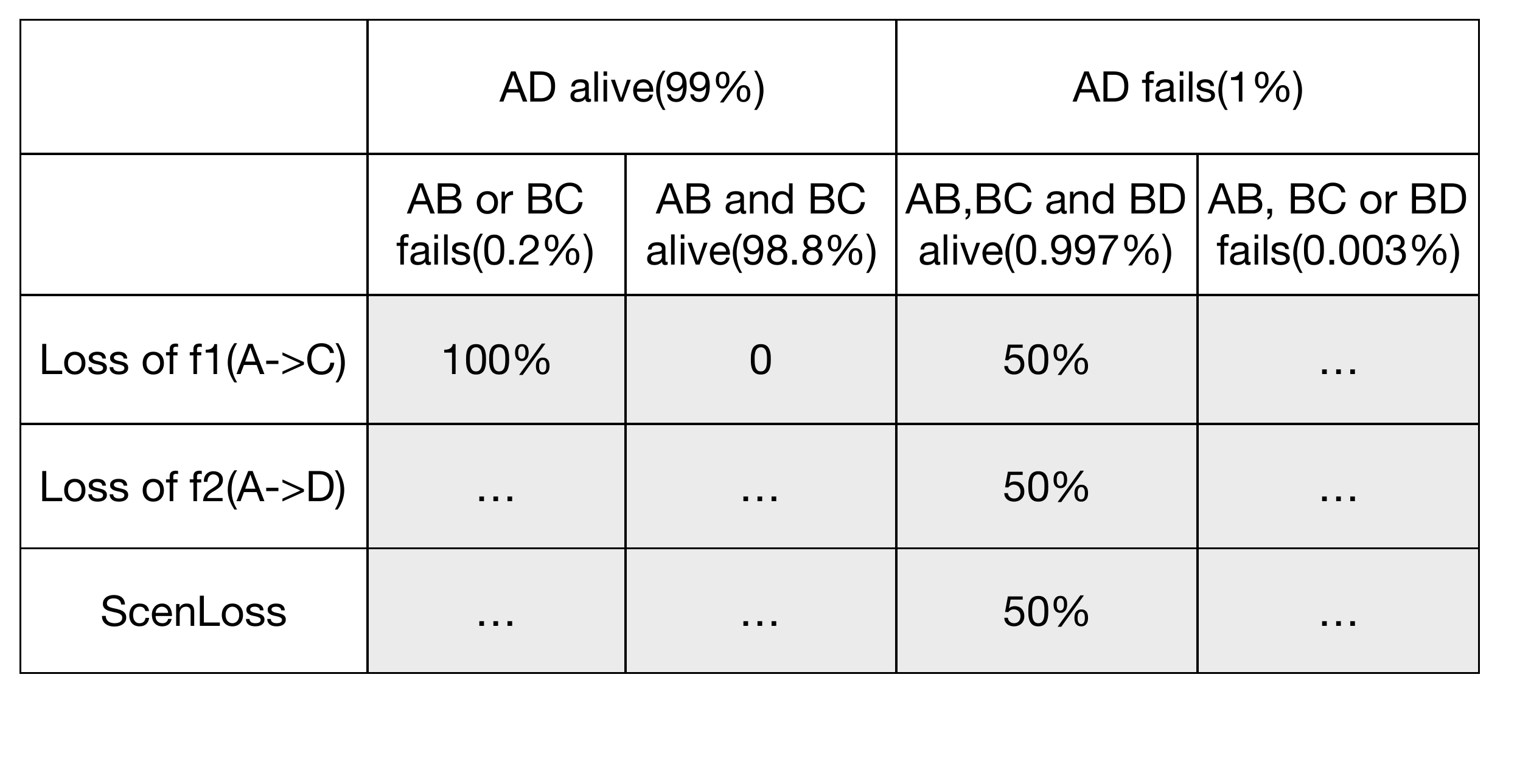}
        \label{fig:smore_table}
\vspace{-0.3in}
    \caption{Loss incurred with SMORE and Teavar.}
        \label{fig:smore_table}
        \vspace{-0.15in}
\end{figure}

\subsection{Illustrating opportunity with \System{}}
Unlike existing TE schemes that work with coarser traffic abstractions, and focus on optimizing an entire traffic aggregate, \System{} abstracts traffic and performance requirements at the granularity of flows. This provides opportunities not available
with existing TE abstractions.

Consider the task of ensuring that no flow sees loss $99\%$ of the time. With existing TE schemes, including Teavar, we need to compute the \ScenLoss{} for each flow and ensure the \ScenLoss{} is $0$ for scenarios that occur $99\%$ of the time. Effectively, this means that the network must be able to carry the traffic of \textit{all flows} in failure scenarios that occur $99\%$ of the time. While this condition is \textit{sufficient} to ensure the flow requirements are met, it is \textit{not necessary}. Specifically, it may be feasible to meet each flow's requirement through a \textit{different combination of failure states}. This provides significant flexibility, and an opportunity to meet a network's bandwidth requirements even when state-of-the-art TE schemes cannot. This is the opportunity that \System{} exploits.


To concretely illustrate the potential opportunity, consider Fig.\ref{fig:flow_example}, where a network must carry
traffic corresponding to a flow $f1$ from source $A$ to destination $C$, and a flow $f2$ from source $A$ to destination $D$.
Consider a requirement that each of $f1$ and $f2$ must support 1 unit of traffic 
$99\%$ of the time. The capacities of the links are shown as in the figure. All links fail with 
probability $0.001$, except link $AD$ which has a failure probability of $0.01$.

\textbf{Can the network meet the bandwidth requirements?}
We first illustrate how the network is able to meet the requirements without any traffic loss. Consider a simple routing strategy of sending $f1$ through $A-B-C$ and always sending $f2$ through $A-D$. Fig.~\ref{fig:optimal_table} shows the loss of each flow over different groups of scenarios using this strategy. For example, the first column represents 
the set of scenarios where link $AD$ is alive, while either link $AB$ or link $BC$ fails. In all these scenarios, we can send 1 unit of traffic for $f2$ without any loss, but cannot send any traffic for $f1$. Consequently, the loss of $f1$ is $100\%$, while the loss of $f2$ is $0\%$. The last row corresponds to \ScenLoss{}, which is $100\%$, since it is the worst loss experienced across all flows in each scenario. 
The red ovals in each row indicate the groups of scenarios where the a particular flow can be sent without loss. Clearly, each of flows $f1$ and $f2$ can be sent without loss in scenarios that occur $99\%$ of the time indicating the bandwidth requirements can be met. Notice however that each flow meets its requirements through a different combination of failure states.
%

\textbf{State-of-the-art TE schemes do not meet the bandwidth requirements.}
We next consider TE strategies such as Teavar~\cite{cvarSigcomm19} and SMORE~\cite{semi_oblivious_nsdi18} 
that minimize \ScenLoss{} in each scenario (or equivalently, minimize MLU, 
or maximize demand scale factor using a maximum concurrent flow formulation).  
Fig. \ref{fig:smore_table} shows the loss of $f1$ and $f2$ in different scenarios
with such scenario-centric TE schemes. The figure shows that $f1$ can only achieve
$50\%$ loss in scenarios that occur $99\%$ of the time.
 
Comparing Fig.~\ref{fig:optimal_table}, and Fig. \ref{fig:smore_table}, a key difference is in the third column, which corresponds to the scenario, where link $AD$ fails, and all other links are alive. 
Here, both Teavar and SMORE route traffic as shown in Fig.\ref{fig:flow_example}. Specifically, $0.5$ units 
is allocated to each of flows $f1$ and  $f2$, translating to an overall loss of $50\%$ for each flow, 
and a \ScenLoss{} of $50\%$. In contrast, to meet the bandwidth requirement, the network must prioritize 
flow $f1$ and serves it without loss for this scenario, even though flow $f2$ sees $100\%$ loss (Fig.~\ref{fig:optimal_table}).
This is acceptable since (i) $f2$ already meets the requirement by performing acceptably in other scenarios; and (ii) it is necessary that $f1$ sees no loss in this scenario to ensure its requirement is met. In doing so, the \ScenLoss{} for this scenario is $100\%$, higher than the $50\%$ achieved by the current TE schemes -- yet this is desirable from the perspective of ensuring the  requirements of all \flow{}s are met.

\textbf{Generalization.}
A simple generalization of this example shows that both Teavar and SMORE may be arbitrarily worse than the network's intrinsic capability. Instead, suppose $A-D$ link has a capacity of $n$, and flow $2$ requires $n$ unit of traffic. The requirement can still be met without any loss following the previous arguments. However, in the scenario where $A-D$ link fails, to minimize \ScenLoss{}, $n/(n+1)$ traffic will be sent from $A$ to $D$, and $1/(n+1)$ traffic will be sent from $A$ to $C$, resulting a loss of $1 - 1/(n+1)$. As $n$ gets larger, the loss will get closer to $100 \%$. 

%% file: design.tex
\section{\System{} design}
\label{sec:design}
We now present \System{}'s design. Given a set of flows and the bandwidth demand associated with the flow, a set of failure scenarios to consider, and the probabilities associated with those scenarios, \System{} decides how to allocate bandwidth to each flow in every failure scenario so the flow sees acceptable performance over scenarios that occur with a desired probability.


A key idea of \System{} is that not all failure scenarios of the network are critical to meeting a particular flow's requirement. Instead, it seeks to determine the set of \textit{critical failure scenarios} associated with each flow, i.e., those failure scenarios where the loss associated with the flow must be acceptable so the objectives can be met.
Further, each flow may be associated with a different set
of critical failure scenarios.

\System{} consists of (i) an offline phase, which involves determining the critical failure scenarios associated with each flow; and (ii) an online phase (executed when a failure occurs) which determines the bandwidth allocation of all flows recognizing that the failure scenario is only critical for a subset of flows identified by the offline phase. 

We start by formalizing \System{}'s performance metric, and discuss \System{}'s offline phase for determining critical failure scenarios.
The offline phase involves solving an Integer Program, which may require substantial computational resources with state-of-the-art solvers. Instead, we present a novel decomposition approach to tackle the problem, and heuristics to aid the offline phase. We then discuss several generalizations related to \System{}.



  

\subsection{Formalizing \System{}'s problem}
Consider a network topology, represented as a graph $G= \langle V, E 
\rangle $. Each link $e \in E$ is associated with a link capacity $c_e$. We use $P$ to represent the set of source-destination pairs. 
Each flow $f\in F$ is associated with traffic demand $d_{f}$ 
that must be sent along the source-destination pair $pr(f)$. 


$Q$ represents the set of failure scenarios. For each scenario $q \in Q$, $p_q$ represents the probability of $q$. Each pair $i$ can use a set of tunnels $R(i)$ to route the traffic. Let $y_{tq}$ represent whether a tunnel $t$ is alive in scenario $q$. We use $x_{tq}$ to denote the bandwidth assigned to tunnel $t$ in scenario $q$, i.e., our designed routing. 
Table~\ref{tab:notations} summarizes notation.




For each flow $f \in F$, we define $\AppLoss{}(f,\beta)$ to be the $\beta^{\text{th}}$ percentile of loss for flow $f$. That is, there 
exist failure scenarios that together occur with probability $\beta$, 
where flow $f$ encounters a loss less than $\AppLoss{}(f,\beta)$.



We start with a formulation where \System{} determines a bandwidth allocation such that the maximum of the $\beta^{th}$ percentile loss across all flows is minimized. Specifically, we consider the following metric that we refer to as \AppOptLoss{} (and may abbreviate as $\alpha$).

%
%
\begin{align}
\AppOptLoss{} (\alpha) = \max_{f \in F} \AppLoss{}(f,\beta)
\end{align}
%
%
This formulation corresponds to a case where \System{} determines a bandwidth allocation such that \textit{all flows} see a loss less than 
$\alpha$ 
over a set of scenarios that occur with probability 
$\beta$. It is easy to generalize \System{} to a context where bandwidth allocations must ensure each flow meets a flow-specific loss threshold (if such an allocation is feasible), as we will discuss in
\S\ref{sec:generalization}.

To ensure each \flow{}'s objectives, 
\System{} must for each \flow{} $f$ select scenarios that together 
occur with probability $\beta$ such that $f$ sees loss less than
$\alpha$ in these scenarios. We denote these scenarios as \textit{critical scenarios} for that flow. We use binary variable $z_{fq}$ to indicate whether scenario $q$ is critical for flow $f$.
If $z_{fq}=1$, $q$ is critical for $f$, and the loss of flow $f$ 
cannot exceed \AppOptLoss{}, i.e., $l_{fq} \le \AppOptLoss{}$.



We next present the formulation below which determines the best routing and choice of critical scenarios that can minimize $\alpha$.
%
\begin{align}
\mbox{($I$)} \ 
& \min_{z,x,l,\alpha} \quad \alpha \nonumber \\ 
\mbox{s.t.} 
& \sum_{q \in Q} z_{fq}p_q \ge \beta \quad \forall f \in F \label{eq:availability}\\
& \alpha \ge l_{fq} - 1 + z_{fq}  \quad \forall f \in F, q \in Q \label{eq:loss_bound} \\
& \sum_{pr(f)=i} (1-l_{fq})d_{f} \le  \sum_{t \in R(i)}x_{tq}y_{tq} \quad \forall i \in P, q \in Q \label{eq:demand} \\
& \sum_{e \in t} x_{tq} \le c_e \quad \forall e \in E,q \in Q \label{eq:capacity}\\
& x_{tq} \ge 0 \quad \forall i \in P, \forall t\in R(i), q\in Q 
\label{eq:non_neg} \\
& z_{fq} \in \{0,1\} \quad \forall f \in F,q \in Q \label{eq:binary} \\
& 0 \le l_{fq} \le 1 \quad \forall f \in F, q \in Q \label{eq:positive_loss} 
\end{align}
In the above formulation, \eqref{eq:availability} ensures that for each flow, we select enough critical scenarios to cover the probability $\beta$. When $z_{fq}=1$, \eqref{eq:loss_bound} becomes $\AppOptLoss{} \ge l_{fq}$ meaning we care about the loss $l_{fq}$. When $z_{fq}=0$, \eqref{eq:loss_bound} is satisfied no matter what $\AppOptLoss{}$ and $l_{fq}$ are, implying we don't care about the loss $l_{fq}$. \eqref{eq:demand} ensures that there is enough bandwidth allocated to each pair. The LHS of \eqref{eq:demand} is the total amount of traffic required to be sent on pair $i$, and the RHS is the total allocated bandwidth on tunnels connecting pair $i$. \eqref{eq:capacity} and \eqref{eq:non_neg} ensure the allocated bandwidth on tunnels will never exceed any link's capacity, and the allocation is non-negative. The final two constraints indicate the $z$ variables are binary, and ensure the loss fractions are between $0$ and $1$.

\begin{table}[t]
    \centering
    \resizebox{0.48\textwidth}{!}{
    \begin{tabular}{|| c || p{6cm}||} 
    \hline
      notation  & meaning\\
    \hline\hline
$Q$ & Set of all scenarios \\
$F$ & Set of flows \\
$P$ & Set of source-destination pairs\\
$R(i)$ & Set of tunnels for pair $i$\\
$E$ & Set of edges\\
$\beta$ & Target probability for which the bandwidth requirement must be met\\
$pr(f)$ & Source-destination pair along which flow $f$ is being sent \\
$d_{f}$ & Traffic demand of flow $f$ \\
$p_q$ & Probability of scenario q\\
$y_{tq}$ & 1 if tunnel $t$ is alive in scenario $q$, 0 otherwise\\
$x_{tq}$ & Allocated bandwidth on tunnel $t$ in scenario $q$(routing variable)\\
$l_{fq}$ & Loss of flow $f$ in scenario $q$\\
$z_{fq}$ & 1 if scenario $q$ is considered for flow $f$ to meet the target probability, 0 otherwise \\
\hline    
    \end{tabular}
    }
    \caption{Notation}
    \label{tab:notations}
    \vspace{-0.35in}
\end{table}



\subsection{Decomposing the problem}
\label{sec:decomposition}
With $z_{fq}$ being a binary variable, (I) is an MIP formulation that simultaneously determines (i) the critical scenarios for each flow ($z_{fq}$ variables); and (ii) how the traffic should be routed in each failure scenario taking into account for which flows that scenario is critical. Solving the MIP (I) can
be challenging for large topology sizes, and as the number of flows increases. 

To tackle this, we draw on the Benders' decomposition~\cite{r2017} algorithm, a systematic way to decompose an optimization problem into two stages, and then iteratively search for the optimal. In this approach, we divide the variables from the original problem into two sets for the two stages. We solve the first-stage problem (the master problem) to get values of variables in the first set. Then with the variables in the first set being fixed, we solve the second stage problem, which can learn constraints to improve the master problem. By adding the constraints learned from the second stage, the master problem can find a better solution of the first set of variables. By repeating these two stages iteratively, the solution gets closer to the optimal.

Fig.~\ref{fig:bender_procedure} shows how we apply the Benders' decomposition algorithm on (I). The master  problem proposes the critical scenarios for each flow, and the subproblem figures out how to route traffic when given a proposed set of critical scenarios for each flow.
In our context, the subproblem itself can be further decomposed into multiple subproblems, one per failure scenario, that each determines a routing for that scenario given information regarding which flow that scenario is critical for. Each of our subproblems is extremely fast to solve, and this allows us to speed up the procedure by solving subproblems in parallel. The second stage subproblems will provide the learned constraints to the master problem so that the master problem can improve its critical scenario proposal in the next iteration.


\begin{figure*}[t]
\begin{minipage}[t]{0.6\textwidth}
    	    \includegraphics[trim={5.8cm 9.8cm 4.3cm 3.8cm}, clip, width=\textwidth]{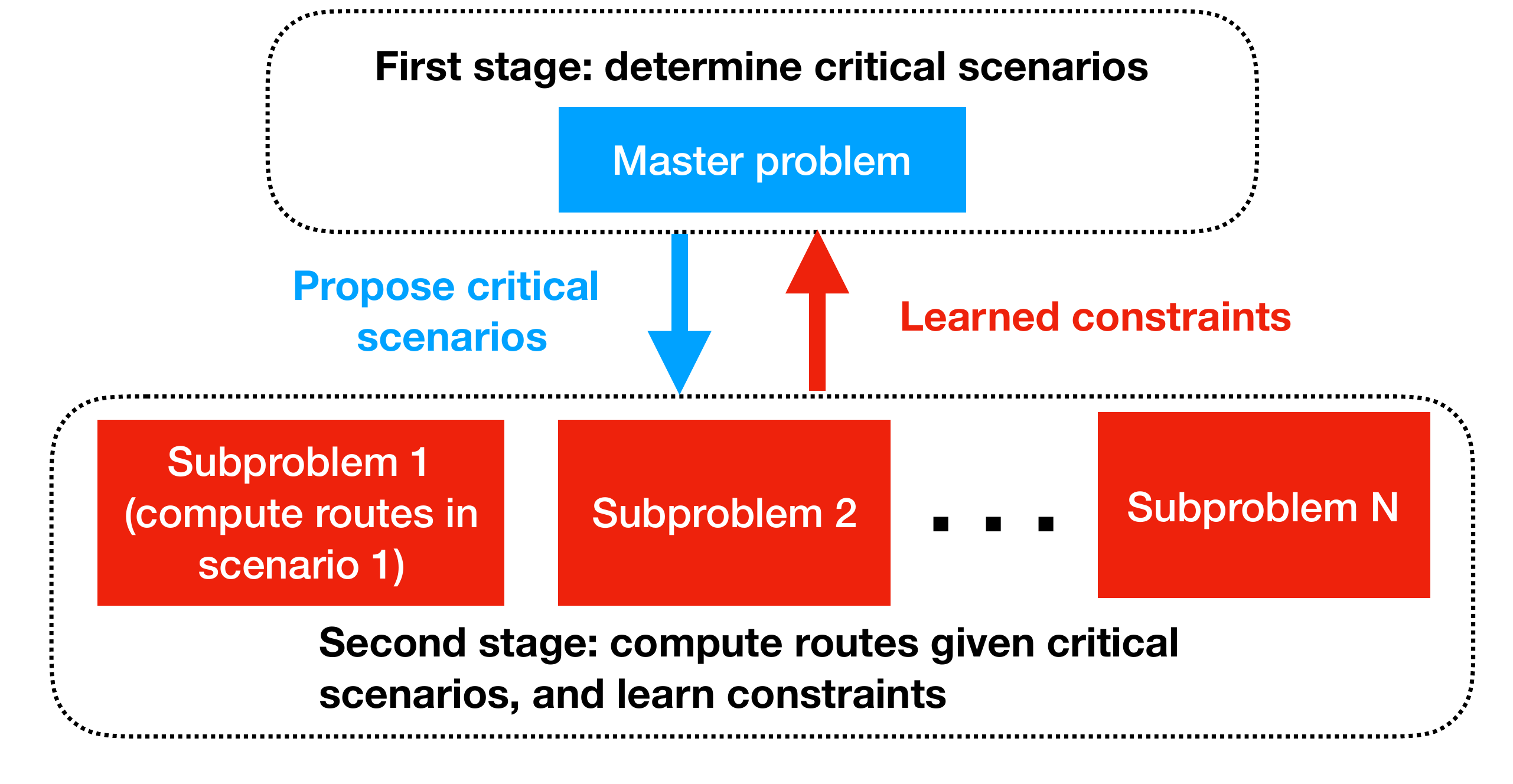}
        \label{fig:bender_procedure}
\vspace{-0.25in}
    \caption{The master problem decides which flows to consider in each scenario, and the subproblems decide how to route in each scenario, and learn constraints to improve the master problem. Note that the subproblems can be solved in parallel. } 
        \label{fig:bender_procedure}
\end{minipage}\hspace{0.1in}
\begin{minipage}[t]{0.35\textwidth}
	    \includegraphics[trim={8.3cm 10.6cm 15.3cm 3.8cm}, clip, width=\textwidth]{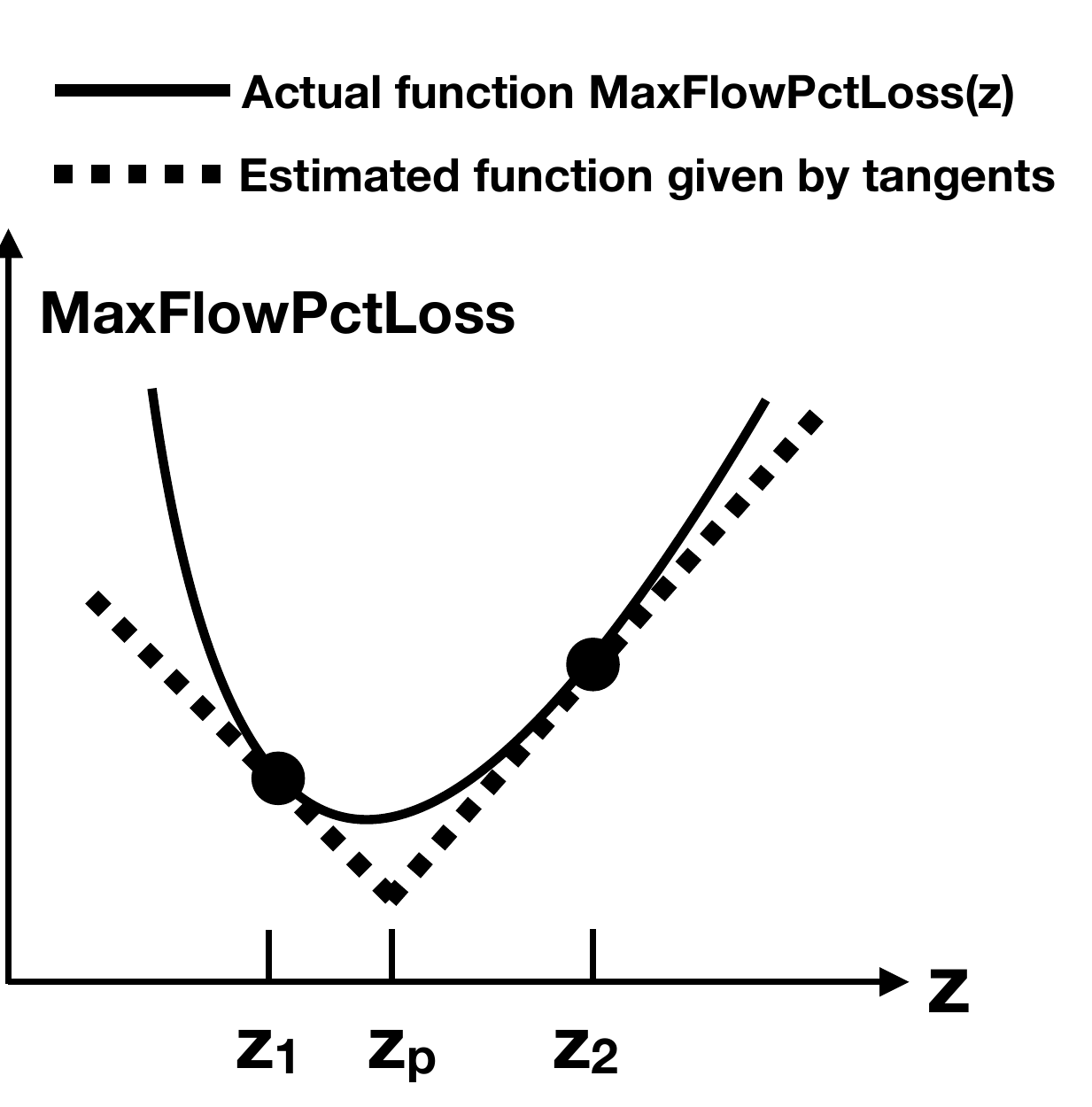}
        \label{fig:z_function}

    \vspace{-0.25in}
    \caption{An example of function \AppOptLoss{}(z). Solving \eqref{eq:inner_min} on $z_1$ and $z_2$ provides two tangents to estimate \AppOptLoss{}(z). And $z_p$ is the minimal based on this estimation.} 
        \label{fig:z_function}
\end{minipage}\hspace{0.1in}
\vspace{-0.15in}
\end{figure*}

Formally, we can rewrite the objective function of (I) as a two step problem by separating the variables into two sets $\{z\}$ (for the first stage) and $\{x, l, \alpha\}$ (for the second stage)
\begin{align}
\min_{z} \min_{x,l,\alpha} \alpha.
\end{align}
The first stage (outer) problem is to decide $z$, which is to determine in each scenario, which flows should be taken care of. With a given $z$, the second stage (inner) problem is to find the routing for each scenario to minimize the loss, which can be seen as a function of $z$. And then (I) can be rewritten as
\begin{align}
\mbox{($I'$)} \ 
& \min_{z} \quad \AppOptLoss{}(z)  \quad
\mbox{s.t.} \eqref{eq:availability}, \eqref{eq:binary} \end{align}
where
\begin{align}
\AppOptLoss{}(z) = \min_{x,l,\alpha}  \alpha  \quad \mbox{s.t.} 
\eqref{eq:loss_bound}, \eqref{eq:demand}, \eqref{eq:capacity} \label{eq:inner_min},\eqref{eq:positive_loss}
\end{align}
We next present some high-level intuition for the inner workings of the decomposition approach (some additional technical clarifications are presented in the Appendix). Fig.~\ref{fig:z_function} shows an example \AppOptLoss{}(z) function. Indeed, the optimal objective value for the inner problem is convex in $z$. This is a well-known property of linear programs. 
%
The decomposition algorithm essentially searches for the minimizer of \AppOptLoss{}(z) iteratively. Although the exact shape of \AppOptLoss{}(z)'s is unknown at any point in the algorithm, solving \eqref{eq:inner_min} gives us one point on the function \AppOptLoss{}(z). Moreover, the dual form of \eqref{eq:inner_min} provides a tangent of \AppOptLoss{}(z).
Thus, we derive an underestimation of \AppOptLoss{}(z) by evaluating it at various $z^i$. Each tangent is a lower bound of function \AppOptLoss{}(z), and the pointwise maximum of these tangents is an underestimate of \AppOptLoss{}(z). Then, we can find the current estimated minimizer of the estimated function(e.g., $z_p$ in Fig.~\ref{fig:z_function}). Solving \eqref{eq:inner_min} at $z^p$ gives a new tangent and a more accurate estimate of \AppOptLoss{}(z). 
The process converges in finite time with an optimal solution (we discuss why in the Appendix).  

The master problem shown below derives an underestimate of the optimal loss using a lower-approximation of \AppOptLoss{} obtained via previously found tangents.
These tangent constraints are learned by solving \eqref{eq:inner_min}, i.e., the second stage problem.
\begin{align}
\mbox{($M$)} \ 
& \min_{z,\AppOptLoss{}} \quad \AppOptLoss{} \nonumber \\ 
\mbox{s.t.} 
& \eqref{eq:availability}, \eqref{eq:binary} \nonumber \\
& \AppOptLoss{} \ge g(z) \quad \forall g \in G \label{eq:cut_} 
\end{align}
$G$ represents the set of all tangents computed so far. So \eqref{eq:cut_} ensures that all tangents are treated as a lower bound of \AppOptLoss{}, forming an estimation.

As mentioned before, \eqref{eq:inner_min} is to find the routing for each scenario, when provided a proposed set of flows for which each scenario is critical. Since the routing in each scenario can be derived independently of that in other scenarios, the problem in the second stage \eqref{eq:inner_min} decomposes by scenarios. Therefore, we solve the following subproblem for each scenario $q\in Q$:
\begin{align}
\mbox{($S_q$)} \ 
& \min_{x,l,\alpha} \quad \alpha \nonumber \\ 
\mbox{s.t.} 
& \alpha \ge l_{fq} - 1 + z_{fq}  \quad \forall f \in F \label{eq:loss_bound__} \\
& 0 \le l_{fq} \le 1 \quad \forall f \in F \label{eq:positive_loss_} \\
& \sum_{pr(f)=i} (1-l_{fq})d_{f} \le  \sum_{t \in R(i)}x_{tq}y_{tq} \quad \forall i \in P \label{eq:demand_} \\
& \sum_{e \in l} x_{tq} \le c_e \quad \forall e \in E. \label{eq:capacity_}
\end{align}
Note that $z_{fq}$ is a parameter in ($S_q$). Suppose the dual variables for \eqref{eq:loss_bound__}, \eqref{eq:positive_loss_}, \eqref{eq:demand_}
 and \eqref{eq:capacity_} are $w_{fq}$, $o_{fq}$, $v_{iq}$ and $u_{eq}$ respectively. Then we will have the  tangent on \AppOptLoss{} as the following function on $z$.
 
 \begin{align}
 g(z) = \sum_{f}(z_{fq}-1)w_{fq} + \sum_{f}o_{fq} + \sum_{i,pr(f)=i}v_{iq}d_{f} + \sum_{e}u_{eq}c_e. \label{eq:cut_function}
 \end{align}
 
 \begin{algorithm}
\caption{Bender's decomposition algorithm}
 \label{alg:bender}
\begin{algorithmic}[1]

\Function{$solve\_master$}{$G$} 
    \State Solve ($M$) with $G$, and get variable $z$
    \State \Return $z$
\EndFunction
\Function{$solve\_subproblem$}{$z$,$q$}  
    \State Solve ($S_q$) and get routing variable $x$, and dual variables $w_{aq}$, $o_{aq}$, $v_{iq}$ and $u_{eq}$
    \State Construct $g$ as in \eqref{eq:cut_function}
    \State \Return $x$, $g$
\EndFunction
\Function{main}{$max\_iterations$}   
    \State $k \leftarrow 0$ 
    \State $G \leftarrow \emptyset$
    \While{$k < max\_iterations$} 
        \State $z \leftarrow solve\_master(G)$  
          \For{$q \in Q$}
        \State $x_q,g \leftarrow solve\_subproblem(z,q)$
        \State $G.add(g)$
      \EndFor
      $k \leftarrow k+1$ 
    \EndWhile 
    \State \Return $x$
\EndFunction

\end{algorithmic}
\end{algorithm}

Note that $S_q$ is a small LP, and for each $q\in Q$, $S_q$ can be solved independently of one another. This means that we can parallelize the solution of these LPs to speed up the solution process.  Algorithm~\ref{alg:bender} provides the pseudocode (Line 13-15 can be executed in parallel). We remark that each iteration, Bender's algorithm yields a routing strategy and the corresponding 
\AppOptLoss{} can be computed easily by sorting the optimal values for $S_q$ and computing the $\beta^{\text{th}}$ percentile.

\subsection{Heuristics to speed up decomposition}
\label{sec:heuristics}
We discuss heuristics which help the decomposition algorithm converge to good solutions faster. 

\textbf{Identifying a good starting point}. 
It is desirable to start with $z$  close to the real minimizer for the outer problem, so that the algorithm requires fewer iterations to converge. As shown in Algorithm~\ref{alg:bender}, the master problem in first iteration will be solved with $G$ being $\emptyset$, which means that there is no tangent to estimate. This will lead to a random $z$, which can be far away from the real minimizer. A simple heuristic is to add constraints $z_{fq}=1$ in (M) indicating all flows are critical in all scenarios. As a further optimization, we observe that a failure scenario can only be critical for a flow if the flow is connected in that scenario. Thus, we add constraints $z_{fq}=0$ in (M) if flow $f$ is disconnected in scenario $q$, and $z_{fq}=1$ otherwise, and use this as an even better starting point. Under either assumption, we have the proposition below (and defer a proof to the appendix).
\begin{proposition}
\label{prop:better_metric}
At the initial step of our algorithm (prior to any iteration of the master problem), the guarantee from our algorithm is already better than that from TeaVar or SMORE.
\end{proposition}



\textbf{Ensure better stability}. With a proper starting point, we may still end up far away from the minimal due to the current estimation being too coarse(For example, in the first few iterations, we don't have enough tangents). Thus, we want to restrict the step we take when we update $z$. That is, in each iteration, we make sure the new estimated minimizer not too far away from the last iterate. One way to achieve this is to add constraints in (M) to limit the hamming distance between current $z$ variable and $z$ variable achieved from last iteration. We remark that this constraint can hurt the convergence of the algorithm to the optimal solution. To circumvent this issue, the Hamming distance constraint can be relaxed, when no improvement is found.

\textbf{Pruning scenarios}. 
We further accelerate the decomposition strategy by recognizing that not all subproblems need to be solved each iteration. In particular, we prune out \textit{perfect scenarios} where all flows can be simultaneously handled without loss. While we have not implemented, we can also prune scenarios with small losses. 
For details, see  Appendix.

\subsection{\System{} online phase}
The offline phase identifies the critical failure scenarios for each flow, which may be seen as hints regarding which flows to prioritize in the online phase. When a failure occurs, a simple linear program ($S_q$) is solved which determines how to allocate traffic to flows so that the loss of flows for which that scenario is critical are minimized. This is a simple and fast LP, and can be easily be integrated with a system such as \Smore{}. The main new aspects of the LP involve leveraging  "hints" regarding critical flows, and running the allocation at the granularity of flows rather than pairs.

\subsection{Generalizations}
\label{sec:generalization}

\textbf{Flows whose loss requirements must be simultaneously satisfied.}
In practice, a service may depend on multiple related flows which we call a flow set. To support a flow set, we need to send the traffic demand of all flows in it. We can define the loss of such flow set to be the max loss across its flows. Under such definition, for example, a flow set having 5\% loss means its worst flow has a 5\% loss. (I) can be rewritten to support flow sets so that each flow set has low loss for probability of $\beta$. And the same decomposition method can also apply to speeding up solving flow set loss. We show the flow set formulation in Appendix.

\textbf{Different loss thresholds across flows.} Different services may have different criteria for meeting desirable performance. For example, service 1 may allow its flow to suffer 5\% loss while service 2 requires all demand to be sent. Our formulation can be easily adapted for requirements that different flows have different loss thresholds. Suppose for any flow $f\in F$, $th(f)$ is the maximum loss it allows for $\beta^{th}$ percentile loss. Then we can change $\eqref{eq:loss_bound}$ in (I) to  
\begin{align}
\alpha \ge (l_{fq} - th(f)) - 1 + z_{fq}  \quad \forall f \in F, q \in Q \label{eq:loss_bound_th} 
\end{align}
With this change, this constraint will only make $\alpha$ positive when $z_{fq} = 1$ and $l_{fq} > th(f)$, i.e., $q$ is critical for flow $f$, and $f$'s loss in $q$ is exceeding its threshold. Thus, if there is a solution where $\alpha$ (the objective) is non-positive, then we have a routing which satisfies the threshold requirements of all flows. If even in the optimal solution, $\alpha$ is positive, then it is impossible to satisfies all requirements. 

\textbf{Heterogeneous percentiles across flows.} 
The percentile targets may vary across flows. An important flow may require a guarantee 99.99\% of time, while a lower priority flow may only require a guarantee 99\% of the time. Since we allow critical scenarios to vary across flows,  
adapting for such requirements is easy in our formulation. Let $\beta_f$ be the required percentile for flow $f$. We can change \eqref{eq:availability} in (I) to  
\begin{align}
\sum_{q \in Q} z_{fq}p_q \ge \beta_f \quad \forall f \in F \label{eq:avaialbility_mul} 
\end{align}
%
In contrast, this is not easy with current TE schemes since they operate at coarser granularities.

\textbf{Shared Risk Link Groups.} Note that our formulation doesn't assume anything regarding the probability distribution of failures. Thus, \System{} can work with any kinds of failure distribution, such as each link has independent failure probability, or a group of links may fail together. As long as the set of failure scenarios is drawn from the failure probability of interest, \System{} can be applied on the set.

\textbf{Capacity augmentation.} The integer programming problem (I) and its decomposition strategy can be generalized to perform minimum-cost capacity augmentation on the network. In this case, we may require that \AppOptLoss{} is constrained to be below a specified value and minimize $\sum_e w_e \delta_e$, where $\delta_e$ is the added capacity to link $e$, which changes the rhs of \eqref{eq:capacity} to $c_e+\delta_e$, and $w_e$ is the per-unit cost of adding capacity. (If there is a fixed-cost, we can include it by introducing a binary variable $a_e$ which takes value $1$ if link $e$ is augmented, and add $\sum_e f_ea_e$ to the cost. To ensure fixed-cost is charged with any augmentation, we add upper-bounding constraints $0\le \delta_e \le u_ea_e$, where $u_e$ is an upper bound on the augmentation.) The decomposition strategy of 
\S\ref{sec:decomposition} generalizes to this setting where $c_e$ is replaced with $c_e + \delta_e$ in \eqref{eq:cut_function} and this cut now describes a tangent of 
\AppOptLoss{} in the $(z,\delta)$ space. Since, for a given routing, the \AppOptLoss{} is no more than the $\beta^{\text{th}}$ percentile of $\ScenLoss{}_q$, it follows that this approach will lead to less costly augmentation pathways to attain performance guarantee for all flows relative to the traditional scenario-based approach.

%% file: cvar_design.tex
\section{\System{} Vs. CVaR-based methods}
\label{sec:cvar_design} 
\System{} seeks to determine bandwidth allocations that minimize the $\beta^{th}$ percentile of flow losses. Minimizing loss at a given percentile (also referred to as Value at Risk or VaR) is a hard problem, and a standard approach in the optimization community involves approximating the same using an approach called \textit{Conditional Value at Risk (CVaR)}. Rather that minimize the $\beta^{th}$ percentile, a CVaR approach involves minimizing the expected loss of the worst $(100 - \beta)^{th}$ percentile of scenarios.

For example, consider a flow which sees a loss of $0\%$, $5\%$ and $10\%$
in three scenarios that respectively occur with probability $0.9$, $0.09$,
and $0.01$. Then, the $90^{th}$ percentile loss (VaR) is $0\%$, but the CVaR is $5 * 0.09 + 10 * 0.01 = 1.45\%$.

Designing networks for probabilistic requirements is a challenging problem, and has only recently received attention from the networking research community. Teavar~\cite{cvarSigcomm19}, a notable and representative recent work in this space
draws on the notion of CVaR and applies it to traffic engineering.

There are three key differences between Teavar~\cite{cvarSigcomm19}, and \System{}.
First, as discussed in \S\ref{sec:motivation}, Teavar computes the \ScenLoss{},
and considers the $\beta^{th}$ percentile of \ScenLoss{}, unlike \System{} 
which focuses on the $\beta^{th}$ percentile of flows. Second, Teavar further conservatively estimates the $\beta^{th}$ percentile using CVaR. In contrast, 
\System{} directly uses VaR. Third, Teavar assumes when a failure occurs,
traffic on a source destination pair is rescaled so the same proportion is maintained on live tunnels. In contrast, \System{} like SMORE~\cite{semi_oblivious_nsdi18} allows greater flexibility in how traffic is split across tunnels. 

\textbf{Enhancing CVaR schemes.}
To appreciate the benefits of \System{}'s approach, we design two new CVaR-based TE schemes, which may be viewed generalizations of Teavar. These schemes allow us to analyze the advantages of directly considering VaR in \System{}, and decouple these benefits from other benefits of \System{}.
The schemes considered are:

$\bullet$ \textit{\Teavarflow{}.} Here, we use CVaR to approximate the computation of \AppOptLoss{}. Instead of directly computing  $\beta^{th}$ percentile loss for flow $f$, i.e., \AppLoss{}($f$, $\beta$), we use CVaR of flow $f$ (denoted by $CVaR(f, \beta)$) to approximate it. Then we seek to optimize the maximum CVaR of all flows, which we denote as 
$\MaxCVaR{}$. Formally, 
\begin{align}
\MaxCVaR{} = \max_{f \in F} CVaR(f, \beta)
\end{align}.
$\bullet$ \textit{\Smoreflow{}.} This is similar to \Teavarflow{} except that we allow greater flexibility in terms of how traffic may be split across tunnels on failure. 

We develop Linear Programming (LP) models for computing the routing and bandwidth allocations associated with these schemes, which we present in the appendix.

\begin{figure}[t]
	\centering
    	    \includegraphics[trim={6.8cm 12.3cm 7cm 6cm}, clip, width=0.48\textwidth]{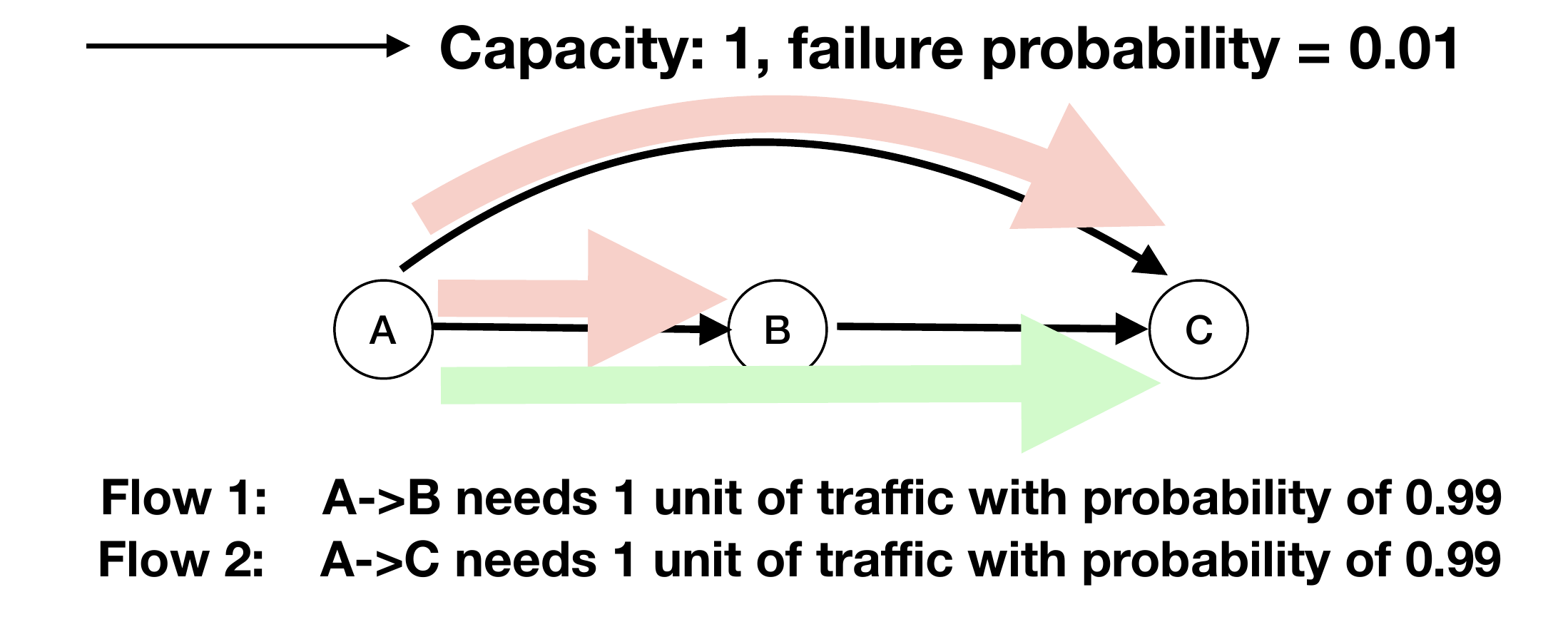}
        \label{fig:cvar_topo}
\vspace{-0.3in}
    \caption{Illustrating why a scheme that optimizes for CVaR may not perform well on loss percentiles.} 
        \label{fig:cvar_topo}
        \vspace{-0.15in}
\end{figure}


\textbf{Illustrating limitations of CVaR.}
We next present an example to show that a scheme that optimizes for CVaR may not perform well when optimizing for the $\%ile$ loss, the measure we truly wish to optimize.

Consider Fig.~\ref{fig:cvar_topo} which shows a topology with two flows $f_1$ and $f_2$. All links have the capacity of 1 and fail with probability $0.01$. Consider a requirement that each of $f_1$ and $f_2$ must support 1 unit of traffic with probability of $0.99$.  

First, observe that $CVaR(f1,0.99) = 100\%$ for any arbitrary routing scheme. This is because flow $f1$ is disconnected in the $1\%$ of scenarios where link AB fails, and the average loss of the worst $1\%$ of scenarios for $f1$ is thus $100\%$. Then, $\MaxCVaR{} = 100\%$ for any arbitrary routing. Thus, a scheme that seeks to optimize $\MaxCVaR{}$ could result in any arbitrary routing.

However, observe that $\AppOptLoss{}$ can be significantly different based on the routing strategy. To see this, consider a first routing strategy that always routes $f1$ along the link AB, and always routes $f2$ along the link AC
(red arrows). Since each link is alive $99\%$ of the time, both $f1$ and $f2$ achieve a $99\%ile$ loss of 0, and $\AppOptLoss{}=0\%$.

In contrast, consider a second routing strategy
which on the failure of link AC, serves $0.5$ units of each flow, and routes flow $f2$ along the green link shown. With this strategy,
the $99\%ile$ loss of $f1$ would be $50\%$.

Thus, a scheme which optimizes $\AppOptLoss{}$
would produce the first strategy, while a
scheme that optimizes $\MaxCVaR{}$ could produce either (or a completely different routing strategy), effectively not performing well in the $\AppOptLoss{}$ metric.

%% file: evaluation.tex
\section{Evaluations}
\label{sec:evaluation}

\begin{figure*}[t]
\begin{minipage}[t]{0.32\textwidth}
	    \includegraphics[ trim={0.3cm 0.2cm 0cm 0cm}, clip, width=\textwidth]{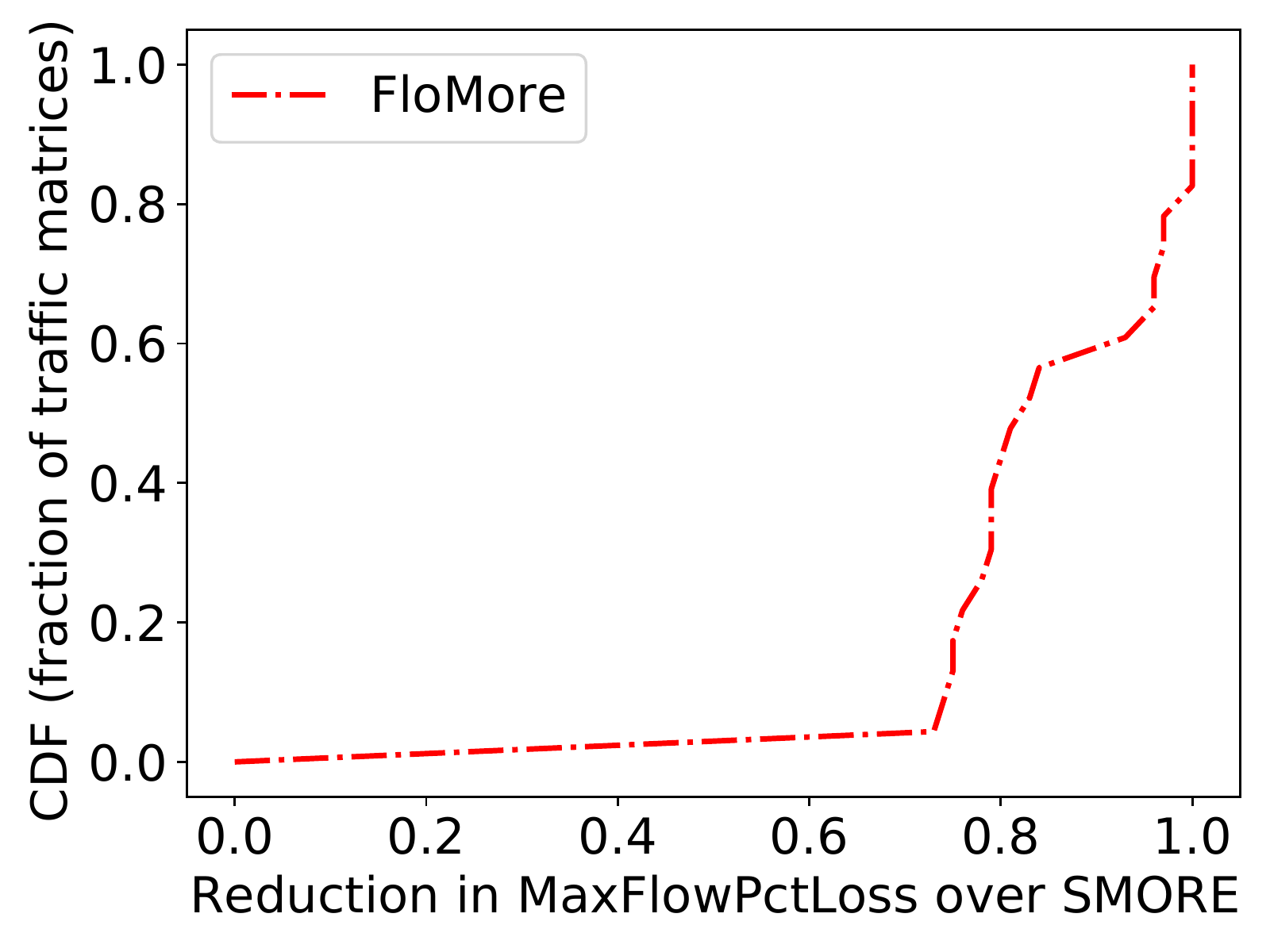}
        \label{fig:beatsmore_cwix}
  \vspace{-0.25in}
    \caption{Benefits over \Smore{} across multiple demands for CWIX topology}
            \label{fig:beatsmore_cwix}
\end{minipage}\hspace{0.1in}
\begin{minipage}[t]{0.32\textwidth}
	    \includegraphics[ trim={0.2cm 0.3cm 0.3cm 0.3cm}, clip, width=\textwidth]{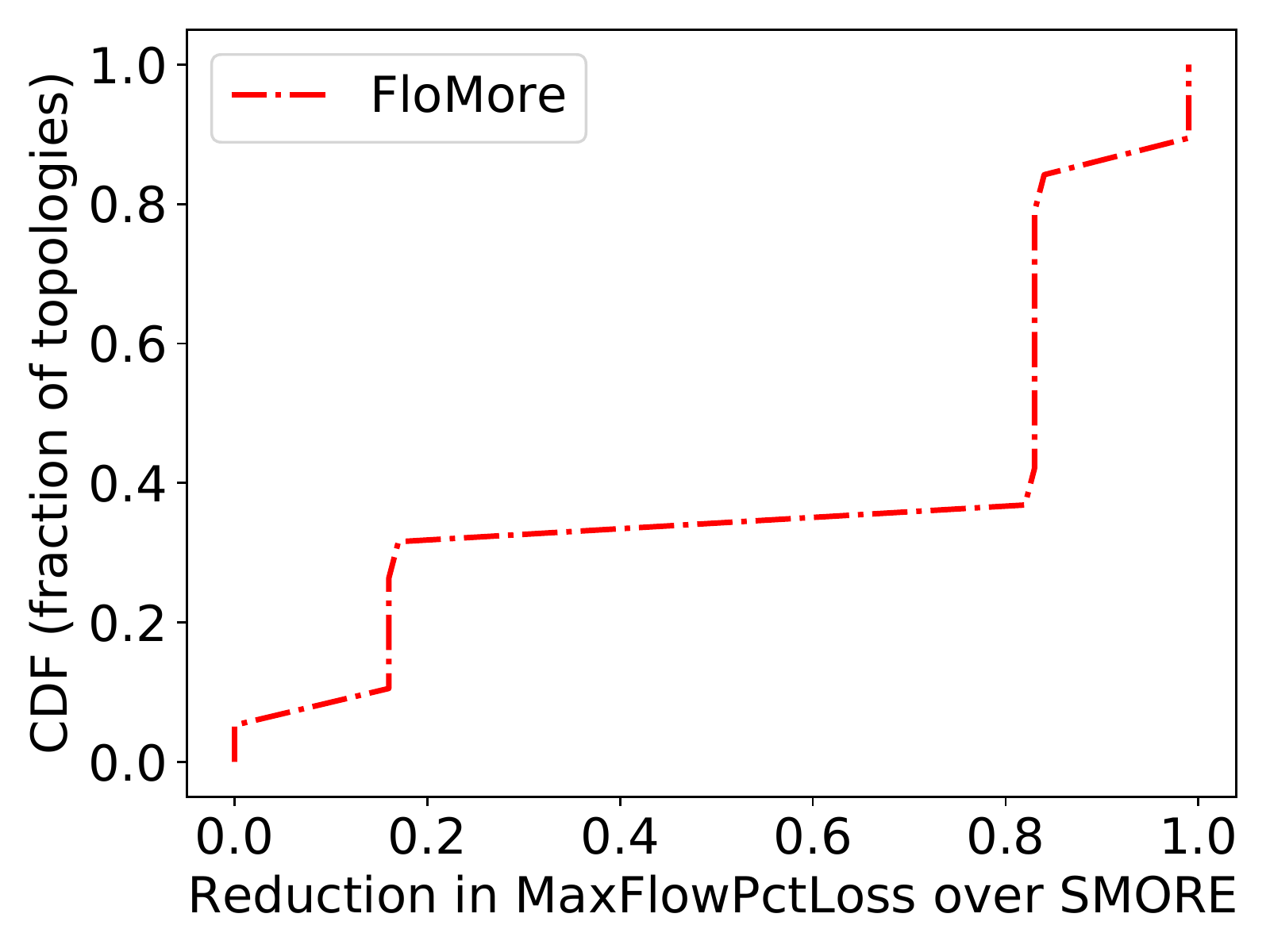}
    \vspace{-0.25in}
        \label{fig:beatsmore}
    \caption{Benefits over \Smore{}}
        \label{fig:beatsmore}
\end{minipage}\hspace{0.1in}
\begin{minipage}[t]{0.32\textwidth}
	    \includegraphics[ trim={1.9cm 3.3cm 4.7cm 2.5cm}, clip, width=\textwidth]{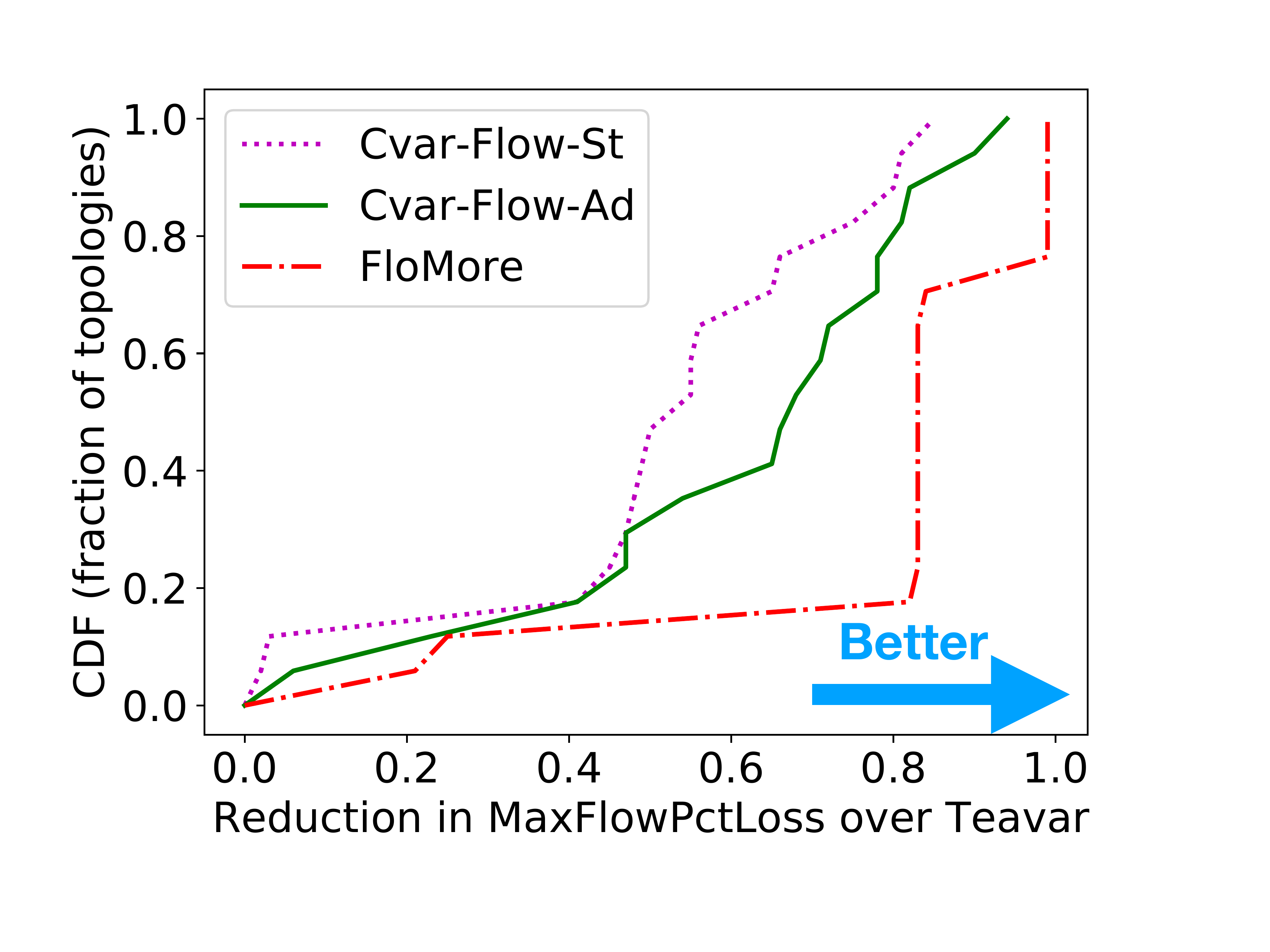}
    \vspace{-0.25in}
        \label{fig:beatteavar}
    \caption{Comparison to \Teavar{} and other CVaR-based schemes}
        \label{fig:beatteavar}
\end{minipage}\hspace{0.1in}
\vspace{-0.3in}
\end{figure*}

\begin{figure*}[t]
\begin{minipage}[t]{0.32\textwidth}
	    \includegraphics[ trim={2.0cm 2.6cm 3.8cm 2.2cm}, clip, width=\textwidth]{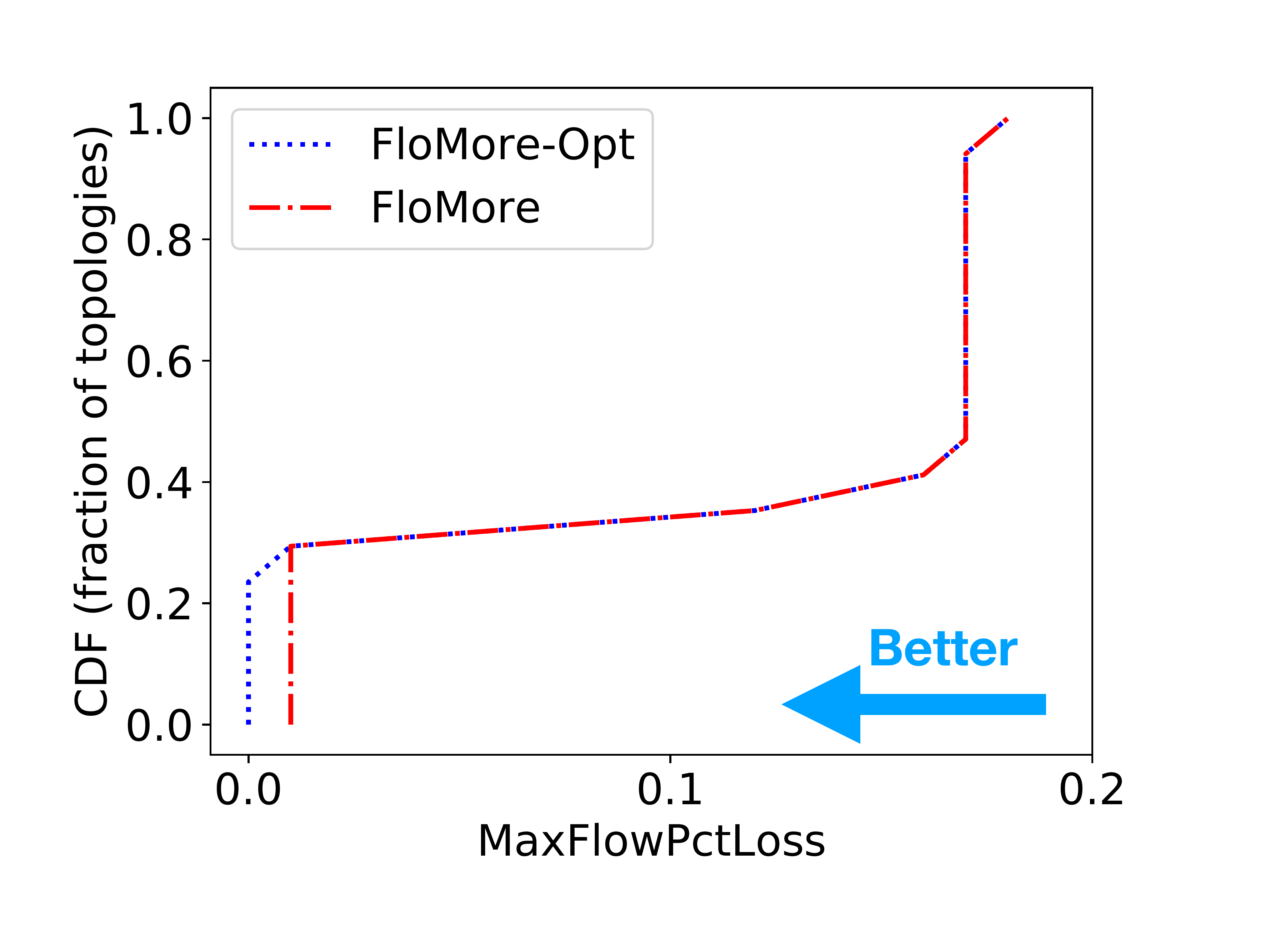}
        \label{fig:bender_vs_ip}

  \vspace{-0.25in}
    \caption{Comparison to optimal}
            \label{fig:bender_vs_ip}
\end{minipage}\hspace{0.1in}
\begin{minipage}[t]{0.32\textwidth}
	    \includegraphics[ trim={1.0cm 0.8cm 0.2cm 0.5cm}, clip, width=\textwidth]{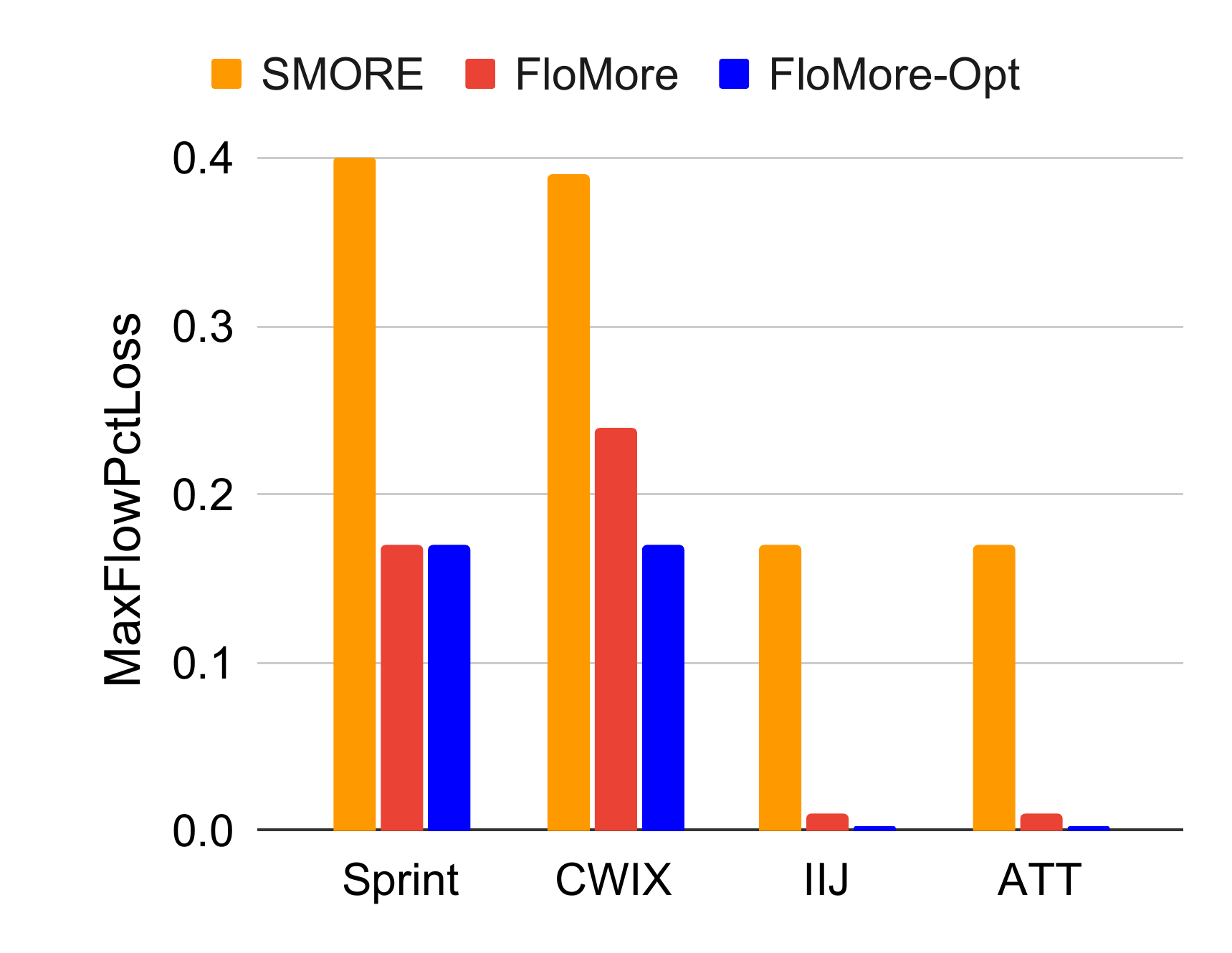}
        \label{fig:beatsmore_k2}

  \vspace{-0.25in}
    \caption{Performance in topologies with richer connectivity}
            \label{fig:beatsmore_k2}
\end{minipage}\hspace{0.1in}
\begin{minipage}[t]{0.32\textwidth}
	    \includegraphics[ trim={1.3cm 2.2cm 0cm 0cm}, clip, width=\textwidth]{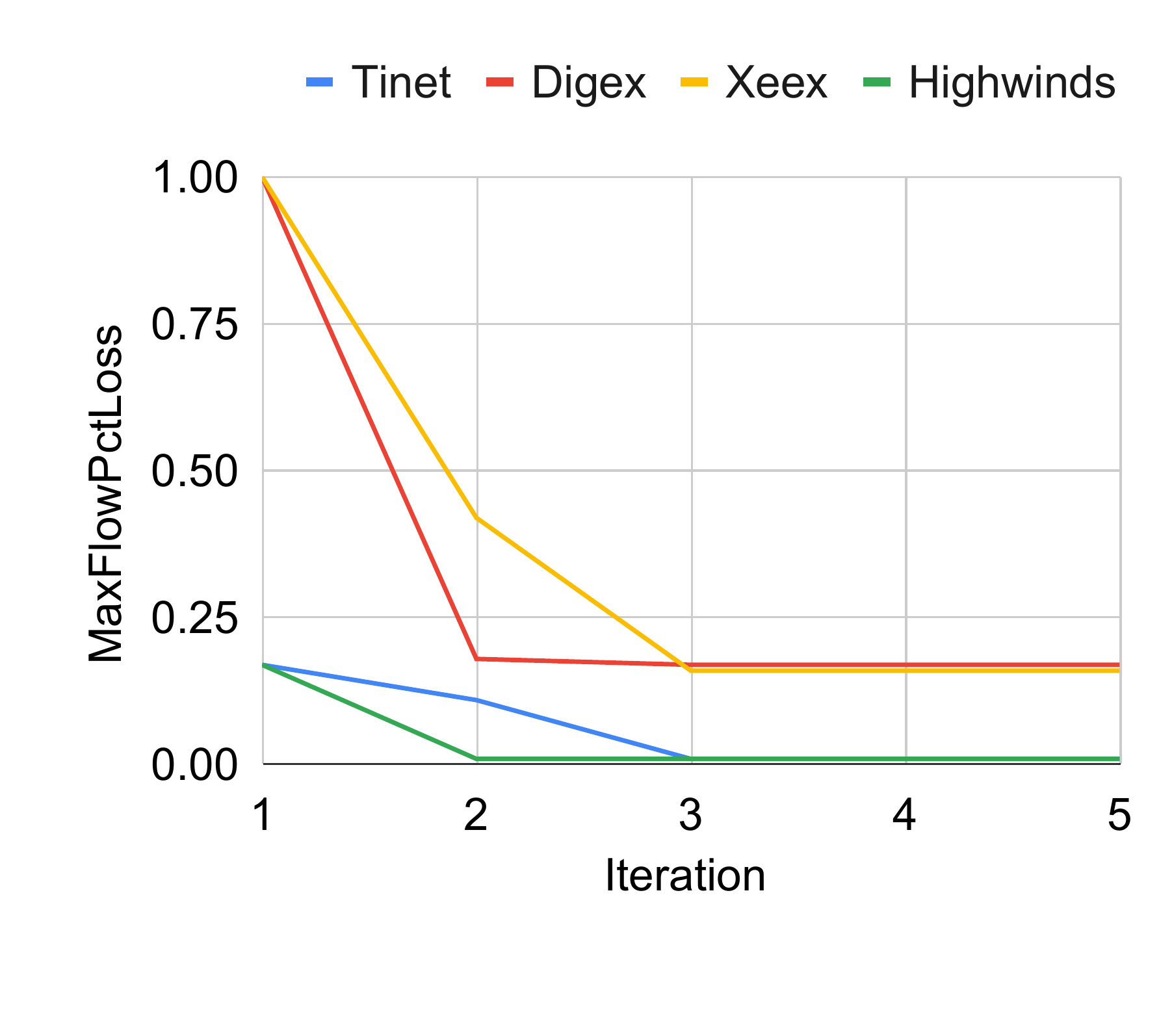}
        \label{fig:loss_change}

  \vspace{-0.25in}
    \caption{Performance improvement with each iteration}
            \label{fig:loss_change}
\end{minipage}\hspace{0.1in}

\vspace{-0.2in}
\end{figure*}

%



We compare \System{}'s performance with 
state-of-the-art TE schemes. We consider the following schemes:

$\bullet$ \textit{\System{} and \SmoreIP{}.}
We primarily report results using \System{}, which uses the decomposition methods and an iterative algorithm introduced in (\S\ref{sec:decomposition}). We set the maximum number of iterations to be 5. To understand how well our schemes perform to optimal,
we also report results with \SmoreIP{} which uses the routing designed by the MIP formulation (I). It is only feasible to run \SmoreIP{} for smaller topologies.  

$\bullet$ \textit{State-of-the-art TE schemes.} We consider \Smore{}, and \Teavar{}, two state-of-the-art TE schemes. \Teavar{} considers the \ScenLoss{} of each scenario (\S\ref{sec:motivation}), and designs a proportional routing scheme that optimizes 
the CVaR of \ScenLoss{}. For each failure scenario, \Smore{} assigns traffic to tunnels in a manner that minimizes the MLU of links, and for a given tunnel selection achieves the optimal MLU (and equivalently, the optimal \ScenLoss{}) for each failure scenario (\S\ref{sec:motivation}). 

$\bullet$ \textit{Enhanced CVaR schemes.} 
We consider two enhanced CVaR schemes 
(\Teavarflow{} and \Smoreflow{}) that we developed (\S\ref{sec:cvar_design}).
While off-the-shelf Teavar uses a proportional routing scheme, and optimizes 
\ScenLoss{}, \Teavarflow{} considers loss percentiles at the granularity of
flows, and \Smoreflow{} additionally considers more flexible routing
(\S\ref{sec:cvar_design}). The purpose of considering these schemes is to decouple \System{}'s benefits owing to its directly optimizing loss percentiles (rather than conservatively estimating using the CVaR measure), from its benefits related to considering losses at a flow granularity.
 
\textbf{Performance metric.} Our primary performance metric for all schemes is 
the \AppOptLoss{} achieved by the scheme (i.e., we consider the $\beta^{th}$ percentile of loss of each flow, and take the maximum across flows).
Note that \Smore{} and \Teavar{} optimize the \ScenLoss{}, but for fairness,
we report the \AppOptLoss{} achieved by these schemes (for any scheme, \AppOptLoss{} is no higher than \ScenLoss{}) 
We evaluate all the schemes based on post-analysis. For each scheme, we determine the routing and bandwidth allocation in each failure scenario,
compute the loss of each flow in each scenario, and then compute \AppOptLoss{}.




\textbf{Topologies and traffic model.}
We evaluate the models above models on 20 topologies obtained from ~\cite{InternetTopologyZoo} and ~\cite{semi_oblivious_nsdi18} (see Table~\ref{tab:topologies} in the Appendix). Our largest network contains 151 edges and 103 nodes. We remove one-degree nodes in the topologies recursively so that the networks are not disconnected with any single link failure. For each node pair in the network, we generate 3 physical tunnels so that they are as disjoint as possible, preferring shorter ones when there are multiple choices. We use the gravity model~\cite{gravity_model} to generate traffic matrices with the utilization of the most congested link (MLU) in the range [0.5, 0.7] across the topologies. 

Although \System{} can model settings with multiple flows per pair, we primarily report results in settings where traffic corresponding to each source-destination pair is modeled as a single flow. We note that \System{}'s benefits are guaranteed to improve (or stay the same) with more flows per pair. To stress \System{}, we evaluate its ability to scale to more flows per pair. We implement all our optimization models in Python, and use Gurobi 8.0~\cite{gurobi} to solve them.
\sgr{Oblivious tunnels?}

\textbf{Failure scenarios.}
For each topology, we use the Weibull distribution to generate the failure probability of each link, like prior work~\cite{cvarSigcomm19}. We choose the Weibull parameter so that the median failure probability is approximately 0.001, matching empirical data characterizing failures in wide-area networks~\cite{Sprint,failures:sigcomm2010,failures:sigcomm2011}. Given a set of link failure probabilities, we sample failure scenarios based on the probability of the occurrence. Our evaluations assume independent link failures but \System{}'s approach easily generalizes to shared risk link groups (\S\ref{sec:generalization}).
We discard scenarios with insignificant probability ($<10^{-6}$). 
Our design target is set as the most 9's for which all flows in the network remain connected for the sampled scenarios. For example, if every flow is connected in scenarios with probability greater than $0.999$, while some flow is disconnected in scenarios with probability more than $0.0001$, we design for $0.999$ (since the network will trivially see an \AppOptLoss{} of 1 when designing for 4 9's).


\subsection{Results}



\textbf{Benefits of \System{}.} 
Fig.~\ref{fig:beatsmore_cwix} compares the \AppOptLoss{} of \System{} and \Smore{}
for the CWIX topology and for 24 different traffic demands.  
For each demand, the \AppOptLoss{} is computed for each scheme, and the reduction in \AppOptLoss{} is determined. The graph shows a CDF of the reduction in \AppOptLoss{} relative to \Smore{} across the traffic demands. In most cases, \System{} 
reduces \AppOptLoss{} by more than 70\%. 

Fig.~\ref{fig:beatsmore} shows the reduction in \AppOptLoss{} achieved by
\System{} relative to \Smore{} across all topologies. In most cases, \System{} 
reduces \AppOptLoss{} by over $80\%$. In some extreme cases, \System{} achieves
a reduction of $100\%$.
%
%
To understand this, consider that there are failure scenarios where the network topology may get disconnected. In such cases, schemes such as \Teavar{} and \Smore{} that optimize \ScenLoss{} (maximum loss across all source-destination pairs in a scenario) cannot count that scenario towards meeting the requirement of any flow.
However, a majority of flows may still be connected, and \System{} may still allow that scenario to count towards the requirement of some of these connected flows.
In some cases, the topology was connected less than $99.9\%$ of the time,
and consequently \Smore{} and \Teavar{} could only achieve a $100\%$ loss at the $99.9\%ile$. In contrast, each individual flow could still be connected in scenarios that occur $99.9\%$ of the time or higher, allowing \System{} to achieve a much 
lower loss at the $99.9\%ile$ (in some extreme cases, \System{} could guarantee 0\% loss). We note that \System{} provides benefits even in more richly connected topologies as we will explore later.


\textbf{Comparison to CVaR-based schemes.}
Fig.~\ref{fig:beatteavar} compares \System{} relative to CVaR-based schemes,
including \Teavar{}, and the new CVaR-based schemes that we designed
(\S\ref{sec:cvar_design}). Each curve shows a CDF of the reduction in
\AppOptLoss{} relative to \Teavar{} for a particular scheme across all topologies. We make several points.

First, \System{} (right-most curve) achieves a significant reduction in \AppOptLoss{} relative to \Teavar{}. These benefits accrue owing to \Teavar{}'s (i) focus on \ScenLoss{} rather than flow losses; (ii) use of CVaR to approximate the percentile; and (iii) assumption of proportional routing rather than allowing traffic to be split over tunnels in more flexible fashion. Second, our enhanced scheme \Smoreflow{} (which allows for flexible routing and considers flow losses)
achieves a significant benefit over \Teavar{}, but does not provide as much benefits as \System{}. This is because the benefits are limited by
the use of the CVaR approximation, while \System{} directly optimizes the percentile. Finally, while \Teavarflow{} provides the lowest reduction in
\AppOptLoss{} relative to \Teavar{}, the benefits are still significant
with a reduction of more than $50\%$ in the median case.
This indicates that the approach of considering flow losses that we advocate in this paper offers significant benefits even in the context of limited routing
flexibility and a CVaR approach.



%
%

\textbf{Comparison to optimal.}
Fig.~\ref{fig:bender_vs_ip} compares \System{} and \SmoreIP{} on 18 topologies except the
two largest ones where we could not run \SmoreIP{} to completion.
The curves overlap significantly indicating   that \System{} performs optimally in most cases. 
%
%
Fig.~\ref{fig:loss_change} shows how \AppOptLoss{} changes each iteration for some typical topologies. While we ran upto 5 iterations of the decomposition algorithm, our results showed the algorithm typically achieved the best performance in fewer iterations.



\textbf{Benefits in richly connected topologies.} 
As discussed above, when a topology gets disconnected in a failure scenario, \Smore{},
and \Teavar{} cannot count that scenario 
towards any flow's requirement (\System{}, and our enhanced CVaR schemes do not suffer from this issue). We consider \System{}'s benefits in more richly connected settings, which
we create by assuming each link consists of two sub-links that fail independently. We ensure the topology remains connected in all sampled failure scenarios. 
Fig.\ref{fig:beatsmore_k2} presents bar charts
comparing the \AppOptLoss{} achieved by \Smore{}, \System{}, and \BenderIP{}
for multiple topologies. The results show \System{} continues to provide benefits over \Smore{}, and performs close to \BenderIP{}.



\begin{figure}[t]
	\centering

	    \includegraphics[ trim={0.3cm 0cm 0cm 0cm}, clip, width=0.48\textwidth]{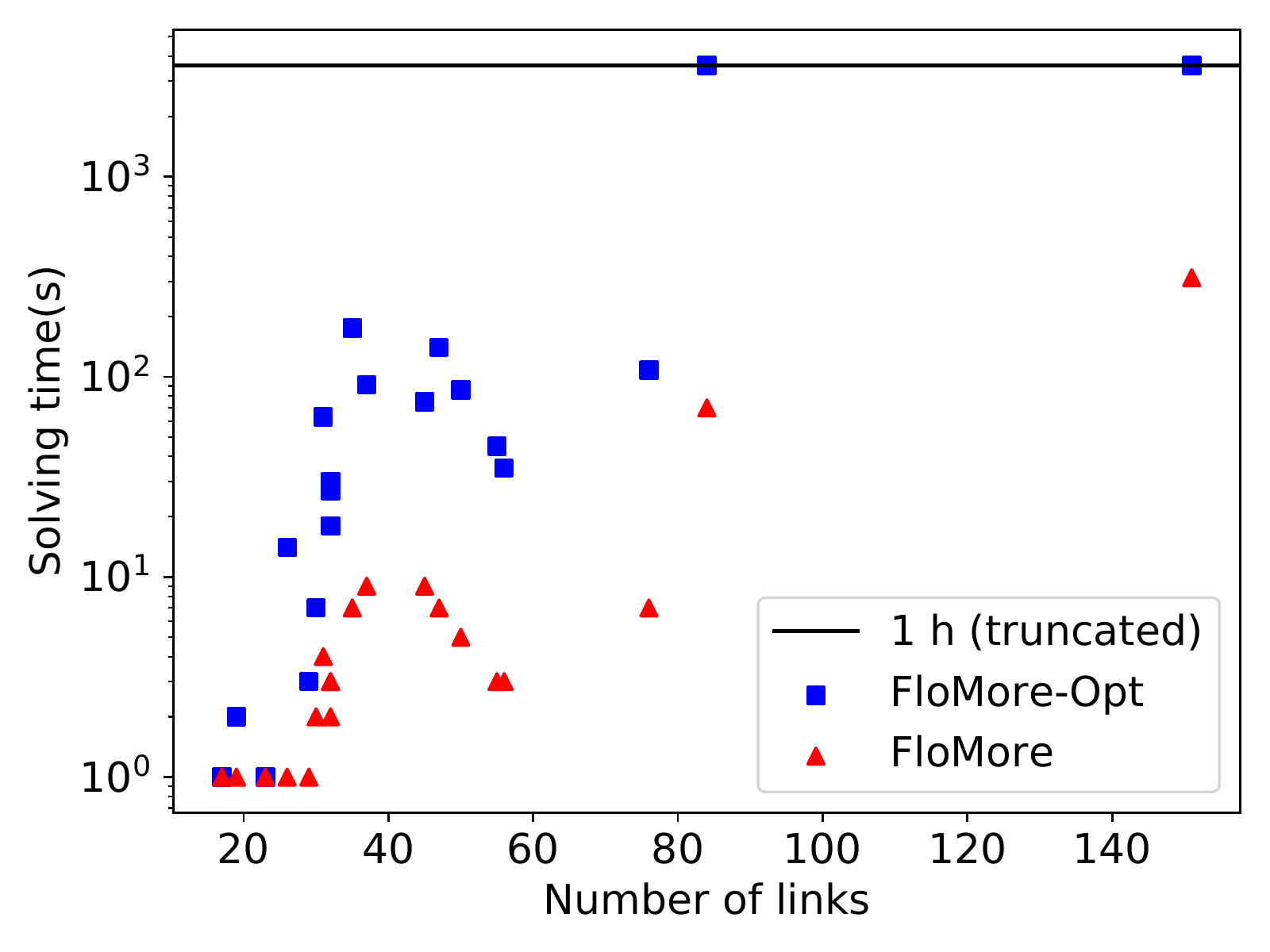}
        \label{fig:time}

  \vspace{-0.3in}
    \caption{Reduction in solving time with \System{}}
    \vspace{-0.2in}
            \label{fig:time}
            
\end{figure}

        

            

\begin{figure}[t]
	\centering

	    \includegraphics[ trim={1.1cm 0cm 2.5cm 0.5cm}, clip, width=0.48\textwidth]{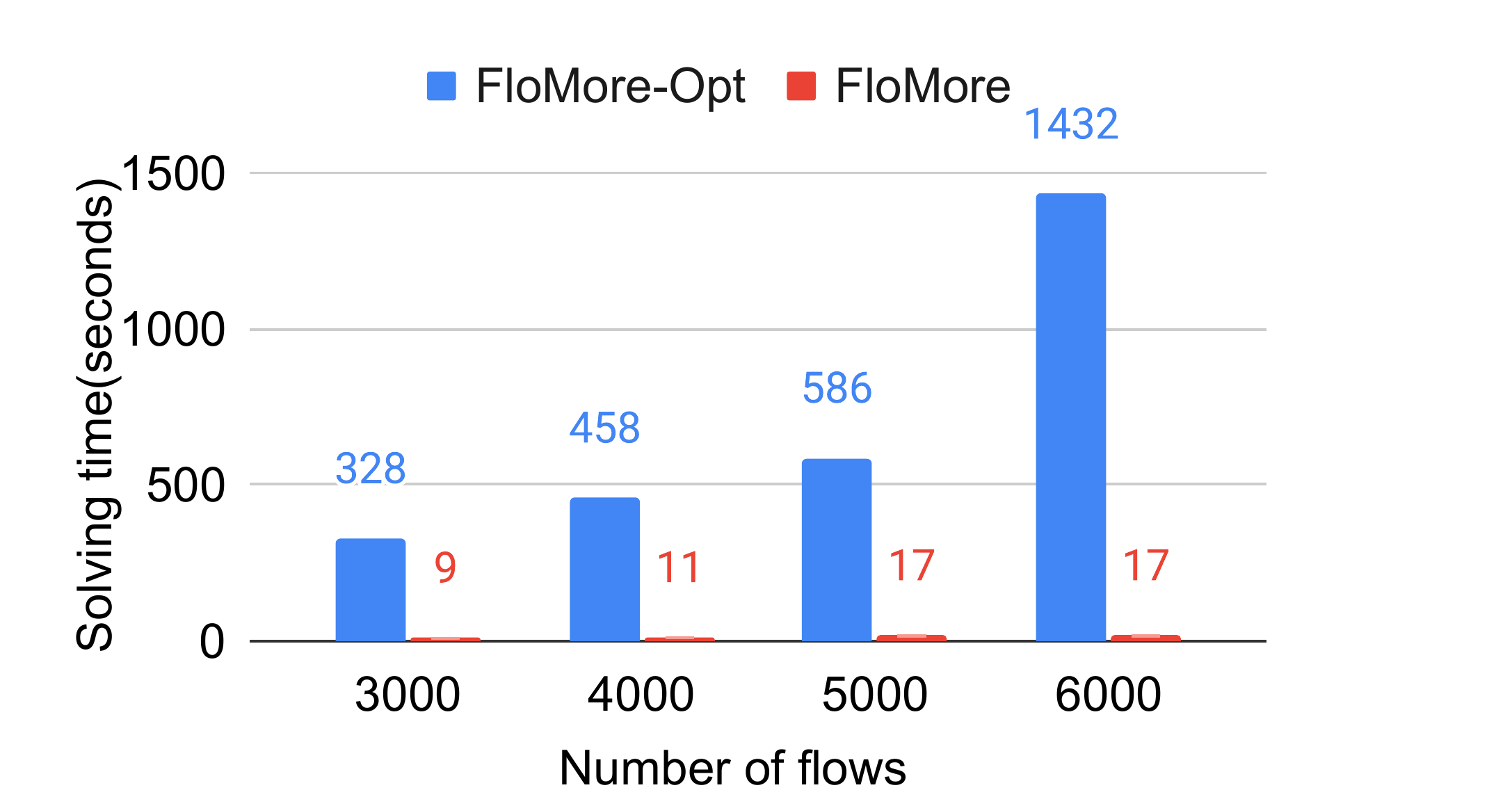}
        \label{fig:time_vs_flows}

  \vspace{-0.25in}
    \caption{Solving time with number of flows}
            \label{fig:time_vs_flows}
            \vspace{-0.2in}
\end{figure}

\textbf{Benefits in computation time.}
Fig.~\ref{fig:time} presents the solving time (Y-Axis) against topology size (X-Axis) for \SmoreIP{} and \BenderIP{}, assuming 5 iterations for \BenderIP{}.
%
Recall that \BenderIP{} solves multiple small LP subproblems in each iteration, and a master problem (a MIP). Even for a relatively large topology like Tinet, solving each subproblem only takes 0.02-0.06 seconds. The master problem is much smaller than the IP (I), and only takes 13-17 seconds for Tinet.
The time we report for \System{} includes the solving time of the master and subproblems in each iteration, and are based on solving upto 10 subproblems in parallel.


Fig.~\ref{fig:time} shows that \BenderIP{} reduces solving time significantly, and is within 10 seconds for most topologies. In the two largest topologies (Tinet and Deltacom), \SmoreIP{} cannot finish within 1 hour, while \BenderIP{} completes 5 iterations for Tinet in 70 seconds, and for Deltacom in around 300 seconds. In both these cases, the \AppOptLoss{} achieved was less than $1\%$ after 2 iterations indicating we could have terminated even earlier.  Several optimizations are possible with our decomposition approach (\S\ref{sec:heuristics}) that we have not explored which can further reduce solving time.

Interestingly, we found that the time to solve CVaR based schemes including Teavar is often significantly higher than \System{}. For example, for HighWinds,  \System{} takes less than 10 seconds while Teavar takes approximately 36 minutes. 
Although CVar based schemes solve a single LP, their solving time can be large since they bundle all the enumerated scenarios in a single problem.




While our results so far assume a single flow per source-destination pair, we next explore \System{}'s scaling with more flows.
Fig.~\ref{fig:time_vs_flows} shows the solving time for \Smore{} and \System{} with the number of flows for the Darkstrand topology, with flows being randomly assigned to source-destination pairs. The solving time of
\SmoreIP{} grows quickly exceeding 20 minutes with 6000 flows, while \System{}'s solving time
grows slowly and is under 20 seconds in all cases.

%% file: relatedwork.tex
\section{Related work}
\label{sec:related}
Researchers have recently started exploring the design of traffic engineering mechanisms with probabilistic requirements in mind~\cite{cvarSigcomm19,lancet:sigmetrics20}. Besides Teavar, Lancet~\cite{lancet:sigmetrics20} 
designs protection routing schemes with the requirement that
a network should not experience congestion over failures scenarios that together
occur with a certain probability. The primary metric in Lancet is MLU, and Lancet only considers a scenario acceptable if the entire traffic matrix can be handled. In contrast, \System{} focuses on losses of flows, and allows for each flow's requirement to be met through a different combination of failure scenarios.

Another recent work NetDice~\cite{netdice:sigcomm20} focuses on verifying that network configurations meet a probabilistic requirement. The focus is on distributed control planes (shortest paths, route redistribution, BGP etc.), and verifying properties such as path length for a given configuration. 
Earlier work~\cite{ProbNetKAT} allows modeling probabilistic network behaviors, such as packet delivery probability on failures. In contrast, \System{} focuses on synthesizing bandwidth allocations so the loss percentile associated with every flow is acceptable, while ensuring link capacity constraints are met in each scenario.

Robust network design has received much attention~\cite{pcf_sigcomm20,ffc_sigcomm14,r3:sigcomm10,DRCN14, nsdiValidation17}. These works guarantee the network remains congestion-free over all failure scenarios with $f$ simultaneous link failures, while allowing
for different levels of flexibility in terms of how networks adapt to failures. All these works capture the performance of any given scenario
using metrics such as MLU, or a demand scale factor (in the context
of a max concurrent flow formulation). \System{} is distinguished by a
focus on design given probability of failures, and its emphasis on metrics that capture the performance of individual flows. Recent work~\cite{akella:pldi20} has explored verification of distributed control planes to ensure load is not violated on failures. Earlier work has explored
robust network design under single link or node failures~\cite{MedhiBook,KodialamRestorationTON08,KodialamRestorationSegmentInfocom16,JenSigmetrics11,applegate:sigmetrics04,zheng_icnp2016,FortzThorup}, and robust design across traffic matrices~\cite{oblivious:sigcomm03,applegate:sigmetrics04,cope:sigcomm06,TrafficMultiMatrix}.

Centralized TE schemes~\cite{b4,swan} optimally route traffic leveraging network-wide views 
balancing between throughput and fairness. Other schemes~\cite{jrex:JSAC13} assign bandwidth to 
flows so overall utility is maximized.
All these works optimize metrics related to 
the entire traffic matrix for any given network snapshot, and do not take
failure probabilities into account. In contrast, \System{} determines
bandwidth allocations by looking across scenarios while considering
failure probabilities, and determines critical flows that must be prioritized
for each scenario.  A linear program is decomposed in ~\cite{jrex:JSAC13} to 
distribute the centralized TE controller, while \System{} involves decomposing an
Integer Program.
Finally, much research explores how to re-route traffic to restore connectivity on failures~\cite{MPLSReroute,IPFastReroute,FCPSigcomm07,Infocom14failureconnectivity,TONProtection11} but does not consider meeting flow
loss percentiles.






%% file: conclusion.tex
\section{Conclusions}
In this paper, we have presented \System{}, a new approach
for designing cloud provider WANs in a manner that meets the bandwidth requirements of flows over failure scenarios that occur with a desired probability. Unlike existing TE schemes 
that seek to meet the requirements of all flows over the same set of failure scenarios making them unduly conservative, \System{} exploits a key opportunity that each flow could meet its bandwidth requirements over a different set of failure scenarios. As part of \System{}, we have presented an approach to optimize the $\beta^{th}$ percentile of bandwidth losses of all flows, a hard problem, and tackled the same using a novel decomposition approach accelerated with problem-specific insights. We have extended a CVaR-based approach to our setting, a well-accepted method to approximating loss percentiles, and shown that it can be conservative. Evaluations over 20 real topologies show the benefits with \System{} are significant.
\System{} (i) reduces \AppOptLoss{} by over $80\%$ relative to \Teavar{} and \Smore{} for over $50\%$ of topologies;
 and (ii) has a solving time of tens of seconds, acceptable for the offline phase, and scales well with the number of flows. Overall, the results show the promise of \System{}.
\textbf{This work does not raise any ethical issues.}

%% file: appendix.tex
\section{appendix}

\textbf{Clarifications regarding Fig.~\ref{fig:z_function}}
The optimal objective value for the inner problem is convex in $z$. This follows from the fact that if, for all $i$,  $(x^i,\text{loss}^i,\alpha^i)$ is feasible when $z=z^i$, then for some multipliers $\lambda_i\ge 0$ such that $\sum_i \lambda_i = 1$, the solution $\sum_i \lambda_i (x^i,\text{loss}^i,\alpha^i)$ is feasible when $z=\sum_i\lambda_i z^i$. This shows that, the optimal value of the inner problem at $z$ is no more than $\sum_i \lambda_i \alpha^i$. The dual form of \eqref{eq:inner_min} provides a tangent of \AppOptLoss{}(z) because its feasible region does not depend on $z$). 

To understand why the process converges in finite time with an optimal solution, consider that the inner problem is always feasible with $(x,\text{loss},\alpha) = (0,1,1)$ and bounded between $0$ and $1$. If we use an extreme point of the dual feasible region to generate a tangent (as is the case using dual simplex algorithm), in finitely many iterations, a cut is developed for each extreme point, and we have an accurate representation of \AppOptLoss{}(z). 

\textbf{Formulation for supporting related flows.} In \S\ref{sec:generalization}, we mention that a set of flows may require their loss requirements to be simultaneously met. Suppose $A$ represents the set of flow sets, and each flow set $a \in A$ contains multiple flows along different source destination pairs. Let $d_{ai}$ be $a$'s required demand along source destination pair $i$. We use $l_{aq}$ to denote the loss of flow set $a$ in scenario $q$, and we define $l_{aq}$ as the max loss across its flows (i.e., at least $(1-l_{aq})d_{ai}$ can be sent along each pair $i$). Then the following formulation is a variant of (I) to minimize the max $\beta^{th}$ percentile loss across flow sets.

\begin{align}
\mbox{($I'$)} \ 
& \min_{z,x,l,\alpha} \quad \alpha \nonumber \\
\mbox{s.t.} 
& \sum_{q \in Q} z_{aq}p_q \ge \beta \quad \forall a \in A \label{eq:availability__}\\
& \alpha \ge l_{aq} - 1 + z_{aq}  \quad \forall a \in A, q \in Q \label{eq:loss_bound___} \\
& \sum_{a} (1-l_{aq})d_{ai} \le  \sum_{t \in R(i)}x_{tq}y_{tq} \quad \forall i \in P, q \in Q \label{eq:demand__} \\
& \sum_{e \in t} x_{tq} \le c_e \quad \forall e \in E,q \in Q \label{eq:capacity__}\\
& x_{tq} \ge 0 \quad \forall i \in P, \forall t\in R(i), q\in Q 
\label{eq:non_neg__} \\
& z_{aq} \in \{0,1\} \quad \forall a \in A,q \in Q \label{eq:binary__} \\
& 0 \le l_{aq} \le 1 \quad \forall a \in A, q \in Q \label{eq:positive_loss__} 
\end{align}

\textbf{Formulations for CVaR-based schemes} The following formulation, \Smoreflow{}, minimizes the maximum conditional value at-risk across all flows. It allows the routing strategy to depend on each scenario. 


\begin{alignat}{2}
&&\min_{x,t,\theta,\alpha,s}\quad & \theta \nonumber\\ 
&&\mbox{s.t.}\quad & \theta \ge \theta_f \quad \forall f \in F \\
&&& \theta_f \ge \alpha_{f} +  \frac{1}{1-\beta}\sum_{q \in Q} p_q s_{fq}  \quad \forall f \in F \\
&&& \alpha_{f} + s_{fq} \ge l_{fq}   \quad \forall f \in F, q \in Q \\
&&& s_{fq} \ge 0 \quad \forall f \in F, q \in Q \\
&&& \eqref{eq:demand}, \eqref{eq:capacity} \nonumber
\end{alignat}

Here, $l_{fq}$ is the loss for flow $f$ in scenario $q$, $\theta_f$ models the conditional value-at-risk for flow $f$, and $\theta$ models $\max_f \theta_f$.

The following formulation, CVar-Flow-St, is derived from CVar-Flow-Ad by requiring that the routing strategy is the same across all scenarios, {\it i.e.}, we add the requirement that $x_{tq} = x_t$ for all $q$. More concretely, we obtain:
\begin{alignat}{3}
&&\min_{x,t,\theta,\alpha,s} \quad &\theta \nonumber\\ 
&&\mbox{s.t.} \quad & \theta \ge \theta_a \quad \forall f \in F \\
&&& \theta_f \ge \alpha_{f} +  \frac{1}{1-\beta}\sum_{q \in Q} p_q s_{fq}  \quad \forall f \in F \\
&&& \alpha_{f} + s_{fq} \ge l_{fq}   \quad \forall f \in F, q \in Q \\
&&& s_{fq} \ge 0 \quad \forall f \in F, q \in Q \\
&&& \sum_{pr(f)=i} (1-l_{fq})d_{f} \le  \sum_{t \in R(i)}x_{t}y_{tq} \quad \forall i \in P, q \in Q \label{eq:demandstat} \\
&&& \sum_{e \in t} x_{t} \le c_e \quad \forall e \in E \label{eq:capacitystat}\\
&&& x_{t} \ge 0 \quad \forall i \in P
\end{alignat}

\textbf{Proof of Proposition \ref{prop:better_metric}} 
Let $\alpha_q$ denote the maximum loss across all flows in scenario $q$, {\it i.e.}, $\alpha_q$ is the optimal value of $S_q$ with $z_{aq}=1$ for all $a$. Let $Q'$ be any minimal subset of $Q$ such that $\sum_{q'\in Q'} p_{q'} \ge \beta$ and for $q'\in Q'$ and $q\not\in Q'$, $\alpha_q\ge \alpha_{q'}$. Then, we define $v=\max_{q'\in Q'}\alpha_{q'}$, which is the $\beta^{\text{th}}$ percentile of $(\alpha_q)_{q\in Q}$. In our first step of the algorithm, we set $z_{aq'}=1$ for all $a$ and $q'\in Q'$.
By definition, $\sum_{q\in Q'} p_qz_{aq} \ge \beta$. In particular, for each application $a$ and $q\in Q'$, $l_{aq}\le v$. Therefore, for each $a$, the $\beta^{\text{th}}$ percentile of $l_{aq}\le v$. So, our performance guarantee, which is the maximum across all $a$ of the $\beta^{\text{th}}$ percentile of $l_{aq}$ is no more than $v$. To see that TeaVar guarantees a performance no better than $v$, let $x_{t}$ be the routing strategy obtained using TeaVar and observe that the maximum loss across all flows using $x_t$ for a scenario $q$ is at least $\alpha_q$. Let $r = (1-\beta) - \sum_{q'\not\in Q'} p_q$, $\bar{q}\in Q'$ be any scenario with $\alpha_{\bar{q}} = v$, and $s$ be the corresponding optimal $s_{\bar{q}}$ (in TeaVar formulation). Then, observe that $r\le p_{\bar{q}}$ and $\alpha + s \ge \alpha_{\bar{q}} = v$, where the inequality follows because there is at least one flow with a loss of $\alpha_{\bar{q}}$ since $\alpha_{\bar{q}}$ is the minimum possible loss attainable across all flows for scenario $\bar{q}$. Then, we have that TeaVar objective is no less than  $\alpha + \frac{1}{1-\beta}\sum_{q'\in Q'}p_{q'} s_{q'} +\frac{1}{1-\beta} rs \ge \frac{1}{1-\beta}\bigl(\sum_{q'\in Q'}p_{q'}\alpha_{q'} + rv\bigr) \ge v$, where the first inequality is because $\sum_{q'\in Q'}p_{q'} + r = 1-\beta$, $\alpha+s_{q'} \ge \alpha_{q'}$, and $\alpha+s\ge v$. The second inequality is because $\sum_{q'\in Q'}p_{q'} + r = 1-\beta$ and $\alpha_{q'} \ge v$ for $q'\in Q'$. Moreover, SMORE guarantees a loss of $v$, since the guarantee for flows in any scenario $q'$ not in $Q'$ is $\alpha_{q'}$. It follows that the guarantee from the initial step of our algorithm is at least as good as the one obtained from either SMORE or TeaVar.

\textbf{Topologies summary (\S\ref{sec:evaluation}).} Our evaluation is done using 20 topologies obtained from
~\cite{InternetTopologyZoo} and ~\cite{semi_oblivious_nsdi18}. The number of nodes and the number of edges of each topology is shown in Table~\ref{tab:topologies}.
\begin{table}[t]
    \centering
    \resizebox{0.48\textwidth}{!}{
    \begin{tabular}{||c c c c c c||} 
    \hline
     Topology & \# nodes & \# edges & Topology &  \# nodes & \# edges \\
    \hline\hline
B4 & 12 & 19 & Janet Backbone & 29 & 45 \\
IBM & 17 & 23 & Highwinds & 16 & 29 \\
ATT & 25 & 56 & BTNorthAmerica & 36 & 76 \\
Quest & 19 & 30 & CRLNetwork & 32 & 37 \\
Tinet & 48 & 84 & Darkstrand & 28 & 31 \\
Sprint & 10 & 17 & Integra & 23 & 32 \\
GEANT & 32 & 50 & Xspedius & 33 & 47 \\
Xeex & 22 & 32 & InternetMCI & 18 & 32 \\
CWIX & 21 & 26 & Deltacom & 103 & 151 \\
Digex & 31 & 35 & IIJ & 27 & 55 \\
    \hline
    \end{tabular}
    }
    \caption{Topologies used in evaluation}
    \label{tab:topologies}
    \vspace{-0.2in}
\end{table}

\textbf{More advanced pruning strategies.}
While we have not implemented, we can further accelerate our decomposition scheme by pruning scenarios with small losses. In particular, we may begin with subproblems that yield cuts tight at the master problem solution. Then, we prune subproblems unlikely to have large loss for critical links using memoized feasible solutions for these scenarios found previously. 

In more detail, the main purpose of solving the subproblems is to generate the cuts \eqref{eq:cut_}. Let $\bar{z}$ be the current optimal solution of the master problem. It can be easily argued that we do not need cuts from all subproblems. In fact, a cut from a single subproblem may suffice if after introducing it to the master problem, $\bar{z}$ is no longer optimal. One way this can be guaranteed, at an intermediate iteration, is if we find a subproblem that yields an optimal value higher than the performance guarantee from a routing strategy found in a previous iteration. In fact, this observation suggests a further acceleration strategy, that allows subproblems to be pruned. To see this, observe that the optimal value of $S_{q'}$ can be under and overestimated using previous solutions of $S_{q'}$ for other settings of $z$. Any feasible routing strategy found in a previous iteration for $q'$ can be used to upper-bound the loss for the critical links given by $\bar{z}$. Similarly, a cut from previous iteration, yields an underestimate $g(\bar{z})$. 
Now, we first solve scenario subproblems for which cuts in the master problem are binding at $\bar{z}$ and assume that among these subproblems the largest optimal value at $z=\bar{z}$ is $\bar{\alpha}$. Then, any subproblem for which a previous overestimator (routing strategy) already yields a bound of $\bar{\alpha}$ or less does not need to solved. Since $\bar{\alpha} \ge 0$, the strategy of not solving a subproblem where all flows can be met is a special case of this more general strategy for pruning scenarios.  